\documentclass[aps,prd,twocolumn,a4paper,superscriptaddress]{revtex4-1}
\usepackage{graphicx}
\usepackage{multirow}
\usepackage{latexsym}
\usepackage{hyperref}
\usepackage[british]{babel}
\usepackage{amsmath}
\begin{document}

\title{Revisit of Y-junctions for strings with currents: \\ transonic elastic case}

\author{I. Yu. Rybak}
\email[]{Ivan.Rybak@astro.up.pt}
\affiliation{Centro de Astrof\'{\i}sica da Universidade do Porto, Rua das Estrelas, 4150-762 Porto, Portugal}
\affiliation{Instituto de Astrof\'{\i}sica e Ci\^encias do Espa\c co, CAUP, Rua das Estrelas, 4150-762 Porto, Portugal}

\date{ \today }

\begin{abstract}
We studied the formation of Y-junctions 
for transonic elastic strings. In particular,
using the general solution for these
strings, which is described by left-
and right-moving modes, we obtained the
dynamics of kinks and Y-junctions. 
Considering the
linearized ansatz for straight strings, 
we constructed the parameter region
space for which 
the formation of Y-junctions due 
to strings collisions is allowed.

\end{abstract}
\pacs{98.80.Cq, 11.27.+d, 98.80.Es}
\keywords{Cosmology; Cosmic strings; String collisions, chiral currents}
\maketitle

\section{Introduction}

Cosmic strings are hypothetical 
objects that were originally described 
by Tom Kibble \cite{Kibble}. 
They appear as a prediction of 
numerous models of early universe
\cite{KibbleCite}. 
To highlight some of them it 
is worthwhile to mention brane 
inflation 
\cite{BurgessMajumdarNolteQuevedoRajeshZhang, DvaliKalloshProeyen, PolchinskiCopelandMyers, SarangiTye, FirouzjahiTye,JonesStoicaTye},
supersymmetric grand unified theory
\cite{JeannerotRocherSakellariadou, CuiMartinMorrisseyWells,JeannerotPostma, AchucarroCeli,DavisMajumdar} 
and theories of high energy particle physics
\cite{BRANDENBERGER2,  KawasakiKen'ichiSekiguchi, GorghettoHardyVilladoro, FleuryMoore, GripaiosRandal-Williams}.

Some types of cosmic strings 
allow the existence of bound
states, named as Y-junctions. 
They might appear due to collisions
of distinct strings that form 
trilinear vertices. Y-junctions 
are common for non-Abelian strings
\cite{Spergel}, for Abelian-Higgs 
strings of the I type \cite{JacobsRebbi},
for U(1)$\times$U(1) models with
specific value of parameters
\cite{Saffin} and for cosmic strings
from brane inflation 
(cosmic superstrings)
\cite{PolchinskiCopelandMyers}. 
Using approximation 
that cosmic strings are 
infinitely thin, are described 
by Nambu-Goto action,
it was demonstrated that 
kinematic constraints must be
satisfied in order for the Y-junctions
to be produced 
\cite{CopelandKibbleSteer, CopelandKibbleSteer2, CopelandFirouzjahiKibbleSteer}. 
The result of kinematic constraints 
was confirmed by numerical 
simulations in a framework of 
field theory 
\cite{SalmiAchucarroCopelandKibblePutterSteer, BevisSaffin}. 
The analytic description of 
cosmic strings via Nambu-Goto 
action also sheds light on 
dynamics of Y-junctions. 
In particular, one can estimate 
the average growth/reduction of 
string lengths for multi-tension 
cosmic string networks. 
This phenomenon is crucial 
for understanding the 
evolution of cosmic (super)string
networks 
\cite{AvgoustidisShellard, RybakAvgoustidisMartins3}.

Due to nontrivial interactions of 
fields that form a string core, 
cosmic strings might become 
superconducting \cite{Witten}. 
This situation naturally arises 
for supersymmetric 
\cite{DavisBinétruyDavis, DavisDavisTrodden,DavisDavisTrodden2, Allys, Sakellariadou} 
and some non-Abelian strings 
\cite{Everett, HindmarshRummukainenWeir}. 
To obtain an effective description of 
superconducting cosmic strings, models
for infinitely thin strings were developed 
\cite{Witten, CARTER1989, Peter1992, CarterPeter, CarterPeter2, Carter2000}. 
It was also suggested that some 
macroscopic properties can be captured 
by such current carrying Nambu-Goto strings. 
In particular, the barytropic cosmic 
string model, which also comes out from 
dimensional reduction 
\cite{Nielsen, NielsenOlesen}, 
provides an accurate 
depiction of ``wiggly" (noisy) cosmic 
strings 
\cite{Carter90, Vilenkin, Martin, Carter95}.

This study revisits the problem 
of Y-junctions for  
transonic elastic strings. 
We re-examine the exact solution for these 
strings \cite{Carter90, Carter95}, obtain 
left-/right-moving modes, and in line with 
\cite{CopelandKibbleSteer} we describe
evolution of kinks and Y-junctions.
In addition, we obtain
kinematic 
conditions under which the production of 
Y-junction is possible.
 
The problem of Y-junctions for Nambu-Goto 
current carrying strings was initially 
studied in 
\cite{SteerLilleyYamauchiHiramatsu}. 
The authors developed a covariant formalism 
to investigate under which conditions 
the production of Y-junctions is possible. 
The result of paper
\cite{SteerLilleyYamauchiHiramatsu}
claims that for 
\textit{magnetic} (space-like current) 
and \textit{electric} (time-like current) 
current carrying strings the formation 
of Y-junction is impossible, unless 
the newly formed string is described 
by a more general equation of state.
Detailed 
comparison of our result with
work \cite{SteerLilleyYamauchiHiramatsu}
can be found in appendix \ref{Appendix}.

\section{Solution in Minkowski space for transonic elastic strings} \label{Solution in Minkowski space for transonic elastic strings}

In this section we revisit the exact 
solution for transonic elastic strings, 
originally obtained in 
\cite{Carter90, Carter95}, 
with the method developed in 
\cite{Blanco-PilladoOlumVilenkin}. 
This approach allows us to show that
only elastic and chiral strings lead 
to wave-like equations of motion.

We start consideration from the action
\begin{equation}
\label{Action}
S = - \mu_0 \int f(\kappa) \sqrt{-\gamma} d \sigma d \tau,
\end{equation}
where $\mu_0$ is a constant defined by 
the symmetry breaking scale, 
$\left\lbrace \sigma, \, \tau \right\rbrace$
are coordinates on the string worldsheet 
(Latin indexes ``$a$-$d$" run over 
$0$, $1$) with induced metric
\begin{subequations}
\begin{align}
    \label{IndMetric}
   & \gamma_{ab} \equiv x^{\mu}_{,a} x^{\nu}_{,b} \eta_{\nu \mu} \quad \text{and} \\
   \label{TermsA}
    & \kappa \equiv \varphi_{,a} \varphi_{,b} \gamma^{ab} , \\ 
     \label{TermsB}
     & \gamma \equiv \frac{1}{2} \varepsilon^{ac} \varepsilon^{bd} \gamma_{ab} \gamma_{cd} , 
\end{align}
\end{subequations}
$\varepsilon^{ac}$ is the Levi-Civita 
symbol, $\eta^{\mu \nu}$ is Minkowski 
metric (Greek indexes run over 
space-time coordinates $x^{\mu}$ 
from $0$ to $3$), $x^{\mu}_{,a} \equiv \frac{\partial x^{\mu}}{\partial \sigma^{a}}$ 
and $\varphi$ is a scalar function on 
the string worldsheet. 
The function $f(\kappa)$ will be 
defined below.

The stress-energy tensor for the action
(\ref{Action}) can be written as
\begin{equation}
\begin{gathered}
\label{Stress-EnergyTensor}
T^{\mu \nu} \equiv -2\frac{\delta S}{\delta g_{\mu \nu}} = \frac{\mu_0}{\sqrt{-g}} \int  \delta^{(4)}(y-x(\sigma)) \\
 \sqrt{-\gamma} \left( U u^{\mu} u^{\nu} - T v^{\mu} v^{\nu} \right) d \sigma d \tau,
\end{gathered}
\end{equation} 
where $u^{\mu}u_{\mu} = 1$ and 
$v^{\mu}v_{\mu} = -1$ are orthonormal 
timelike and spacelike vectors. 
Mass per unity length $U$ and 
tension $T$ in 
(\ref{Stress-EnergyTensor}) 
are given by expressions  
\begin{equation}
\begin{gathered}
\label{UTcurrent}
U =  f - 2 \kappa f^{\prime}_{\kappa} \Theta\left[ - \kappa f^{\prime}_{\kappa} \right] ,  \\
T =  f - 2 \kappa f^{\prime}_{\kappa} \Theta\left[ \kappa f^{\prime}_{\kappa} \right] , 
\end{gathered}
\end{equation}
where $\Theta[...]$ is a Heaviside 
function and $f^{\prime}_{\kappa} = \frac{\partial f}{\partial \kappa} $ 
(for more details about the 
stress-energy tensor see section 
4 in \cite{Carter2} or 
alternatively section 2 in 
\cite{RybakAvgoustidisMartins}).

We can introduce the speed of 
``wiggles" $c_E$ 
(propagation of transverse perturbations) 
and ``woggles" $c_{L}$ 
(propagation of longitudinal perturbations)
acording to \cite{CarterPeter, Carter95}
\begin{equation}
\label{speeds}
c_E^2 = \frac{T}{U}, \qquad c_L^2 = - \frac{d T}{d U}.
\end{equation}

In this way, for the standard Nambu-Goto 
string both propagations have the speed 
of light, 
$c_E = c_L = 1$. 
It is anticipated to have supersonic 
strings ($c_E > c_L$) for most of regimes 
of superconducting strings 
\cite{Peter1992, Peter1993}. 
Meanwhile, the transonic model 
\begin{equation}
\label{Transonic}
c_L = c_E \leq 1
\end{equation} 
can be considered as an effective 
description of wiggly strings 
\cite{Martin, Carter95} 
and some particular limits of 
superconducting strings 
(see sections 5.8, 5.9 in 
\cite{Carter2}).

Using (\ref{UTcurrent}), the explicit 
form of (\ref{speeds}) can be written 
as
\begin{equation}
\begin{gathered}
\label{speeds2}
 c_E^2 = \frac{f-2 \kappa f^{\prime}_{\kappa} \Theta[\kappa f^{\prime}_{\kappa}] }{f-2 \kappa f^{\prime}_{\kappa} \Theta[- \kappa f^{\prime}_{\kappa}]}, \\
 c_L^2 = - \frac{f^{\prime}_{\kappa}-2 ( f^{\prime}_{\kappa} + \kappa f^{\prime \prime}_{\kappa \kappa}) \Theta[ \kappa f^{\prime}_{\kappa} ] }{f^{\prime}_{\kappa}-2 ( f^{\prime}_{\kappa} + \kappa f^{\prime \prime}_{\kappa \kappa}) \Theta [- \kappa f^{\prime}_{\kappa}]}.
\end{gathered}
\end{equation} 

Substituting (\ref{speeds2}) into 
condition (\ref{Transonic}) for 
transonic strings, one can obtain 
the equation for  $f(\kappa)$
\begin{equation}
\label{FunctionF}
(f^{\prime}_{\kappa})^2 + f f^{\prime \prime}_{\kappa \kappa}=0 \; \Rightarrow \; f=\sqrt{c_1 \kappa + c_2},
\end{equation} 
where $c_1$ and $c_2$ are 
constants of integration.

One can write down the equation of 
state for transonic strings using 
the expressions (\ref{UTcurrent}) 
together with (\ref{FunctionF}) 
\begin{equation}
\label{UT}
U T = f \left( f - 2 \kappa f^{\prime}_{\kappa} \right) = c_2 = m^2,
\end{equation} 
where $m$ is a mass dimensional constant.

We can define $c_1 = \pm m^2$ and 
absorb $m^2$ into the definition of 
$\mu_0$. These manipulations allow 
us to establish the function 
$f(\kappa)$ for transonic elastic 
strings in the following form, 
as also presented in 
\cite{Carter90, Carter95},
\begin{equation}
\label{F}
f(\kappa) =  \sqrt{1 - \kappa}, \quad \kappa \in (-\infty,1], \quad UT = 1.
\end{equation} 

It is known that the transonic model 
has the general wave-like solution 
\cite{Carter90}. 
Let's use the method from 
\cite{Blanco-PilladoOlumVilenkin} to demonstrate that there are only 
two types of strings, 
whose equations of motion can 
be reduced to the wave equation: \textit{chiral} 
(see \cite{CarterPeter2, DavisKibblePicklesSteer}) 
and \textit{transonic elastic} strings. 
We start consideration by writing 
down the equations of motion for 
the action (\ref{Action}) in 
Minkowski space  \cite{RybakAvgoustidisMartins2}
\begin{subequations}
\label{EqOfMot11}
\begin{align}
   \label{EqOfMotBPVOA}
    & \quad \; \; \partial_{a} \left[ \mathcal{T}^{ab}  x^{\mu}_{,b} \right] = 0 ,\\ 
     \label{EqOfMotBPVOB}
    & \partial_{a} \left[ \sqrt{-\gamma} \gamma^{ab} f^{\prime}_{\kappa} \varphi_{,b} \right] = 0,
\end{align}
\end{subequations}
where
\begin{equation}
\begin{gathered}
\label{T}
\mathcal{T}^{ab}  = \sqrt{-\gamma} \left( \gamma^{ab} f - 2 f^{\prime}_{\kappa} \gamma^{ac} \gamma^{bd} \varphi_{,c} \varphi_{,d}  \right) = \\
=  \sqrt{-\gamma} \left( \gamma^{ab} f + \theta^{ab} \right)
\end{gathered}
\end{equation}
(notice the change of the sign in (\ref{T}) due to misprint in equation (6) of \cite{RybakAvgoustidisMartins2}).

Parametrization invariance of the 
string worldsheet allows us to make 
the transformation
\begin{equation}
\label{Transform}
 \mathcal{T}^{ab} \longmapsto \eta^{ab},
\end{equation}
if their determinants are equal 
\cite{Blanco-PilladoOlumVilenkin}
\begin{equation}
\text{det} \mathcal{T}^{ab} = \text{det} \eta^{ab} = -1.
\end{equation}
Let's expand the determinant of 
$\mathcal{T}^{ab}$
\begin{equation}
\label{Det}
\text{det} \mathcal{T}^{a b} = -f^2 -f \, \text{Tr} \theta^a_c - \text{det} \theta^a_c.
\end{equation}
It is easy to check that 
$\text{det} \theta^a_c=0$, 
hence, we are left only with
\begin{equation}
\begin{gathered}
\label{Det2}
\text{det} \mathcal{T}^{ab} = -f^2 + 2 f f^{\prime}_{\kappa} \, \text{Tr} \left[ \gamma^{ac} \gamma^{bd} \varphi_{,c} \varphi_{,d} \right] = \\
=-f \left( f - 2  f^{\prime}_{\kappa} \kappa \right)  = - U T.
\end{gathered}
\end{equation}

It is seen from (\ref{Det2}) that the 
transformation (\ref{Transform}) is 
possible due to parametrization invariance 
only if the function $f(\kappa)$ is defined 
as for transonic elastic strings 
(\ref{F}), or $f(\kappa)$ is defined as 
for chiral strings, where the current 
is a null vector $\kappa \rightarrow 0$ 
\cite{RybakAvgoustidisMartins2}.

The relation (\ref{Det2}) together 
with (\ref{F}) guarantees that the 
equations of motion for the string 
worldsheet (\ref{EqOfMotBPVOA}) has 
the general wave-like solution. 
Choosing the gauge where the 
worldsheet coordinate $\tau$ 
coincides with physical time $t$, 
one can write down the solution for 
(\ref{EqOfMotBPVOA}) in the form of 
left- and right-moving modes
\begin{equation}
\label{StrSolution}
x^0 = \tau, \quad \textbf{x} = \frac{1}{2} \left( \textbf{a}(\sigma_+) + \textbf{b}(\sigma_-) \right),
\end{equation}
where $\sigma_+ = \tau+\sigma$ and 
$\sigma_- = \tau - \sigma$.

Up to this point, we demonstrated how 
to obtain the result of 
\cite{Carter90} in a different manner. 
Let's study the equation of motion for 
the function $\varphi$ (\ref{EqOfMotBPVOB}). 
To do so, we plug 
$f f^{\prime}_{\kappa} = \frac{1}{2}$ 
in (\ref{EqOfMotBPVOB})
\begin{equation}
\begin{gathered}
\label{EqOfMotBPVO2}
\partial_{a} \left[ \sqrt{-\gamma} \gamma^{ab} f^2 f^{\prime}_{\kappa} \varphi_{,b} \right] =  \left[\text{using (\ref{Transform}) } \right] =\\
= \partial_{a} \left[ f \left( \eta^{ab} + 2 f^{\prime}_{\kappa} \sqrt{-\gamma} \gamma^{ac} \gamma^{bd} \varphi_{,c} \varphi_{d}  \right) f^{\prime}_{\kappa} \varphi_{,b} \right] = \\
= \partial_{a} \left[ f f^{\prime}_{\kappa} \left( \eta^{ab} + 2  \kappa \sqrt{-\gamma} \gamma^{ab}  f^{\prime}_{\kappa} \right) \varphi_{,b}  \right] .
\end{gathered}
\end{equation}
Transferring the right-hand side term 
with $\sqrt{-\gamma} \gamma^{ab}$ to the 
left-hand side in (\ref{EqOfMotBPVO2}), 
one obtains
\begin{equation}
\begin{gathered}
\label{EqOfMotBPVO3}  
\partial_{a} \left[ \sqrt{-\gamma} \gamma^{ab} f^{\prime}_{\kappa} f \left( f - 2 \kappa f^{\prime}_{\kappa}   \right) \varphi_{,b} \right] = \\
= \partial_{a} \left[ \sqrt{-\gamma} \gamma^{ab} f^{\prime}_{\kappa} \varphi_{,b} \right] = \partial_{a} \left[ f f^{\prime}_{\kappa} \eta^{ab} \varphi_{,b}  \right] = 0.
\end{gathered}
\end{equation}
Taking out the constant 
$f f^{\prime}_{\kappa}$ from the 
differentiation operation in 
(\ref{EqOfMotBPVO3}), we derive the 
following equation
\begin{equation}
\begin{gathered}
\label{EqOfMotBPVO4}
\partial_{a} \left[ \eta^{ab} \varphi_{,b}  \right] = 0,
\end{gathered}
\end{equation}
which general solution is given by 
\begin{equation}
\label{CurrentSolution}   
\varphi = \frac{1}{2} \left( F(\sigma_+) + G(\sigma_-) \right).
\end{equation}

The normalization of $|\textbf{a}^{\prime}|$ 
and $|\textbf{b}^{\prime}|$ are connected 
with values of the current as
\begin{equation}
\label{Normalization}
\textbf{a}^{\prime \, 2}(\sigma_+) = 1 - F^{\prime \, 2}(\sigma_+), \quad \textbf{b}^{\prime \, 2}(\sigma_-) = 1 - G^{\prime \, 2}(\sigma_-)
\end{equation}
for right- and left-moving modes.

Alternative treatment 
of elastic strings, as a Kaluza-Klein 
projection of standard Nambu-Goto strings
in a space-time of 5-dimensions, can
be found in \cite{CarterSteer}. In 
this approach the relation 
(\ref{Normalization}) can be seen
as a normalization for unity 
of 4-dimensional vectors of
left- and right-moving modes.

\begin{figure}[h]
\centering
		\includegraphics[scale=0.53]{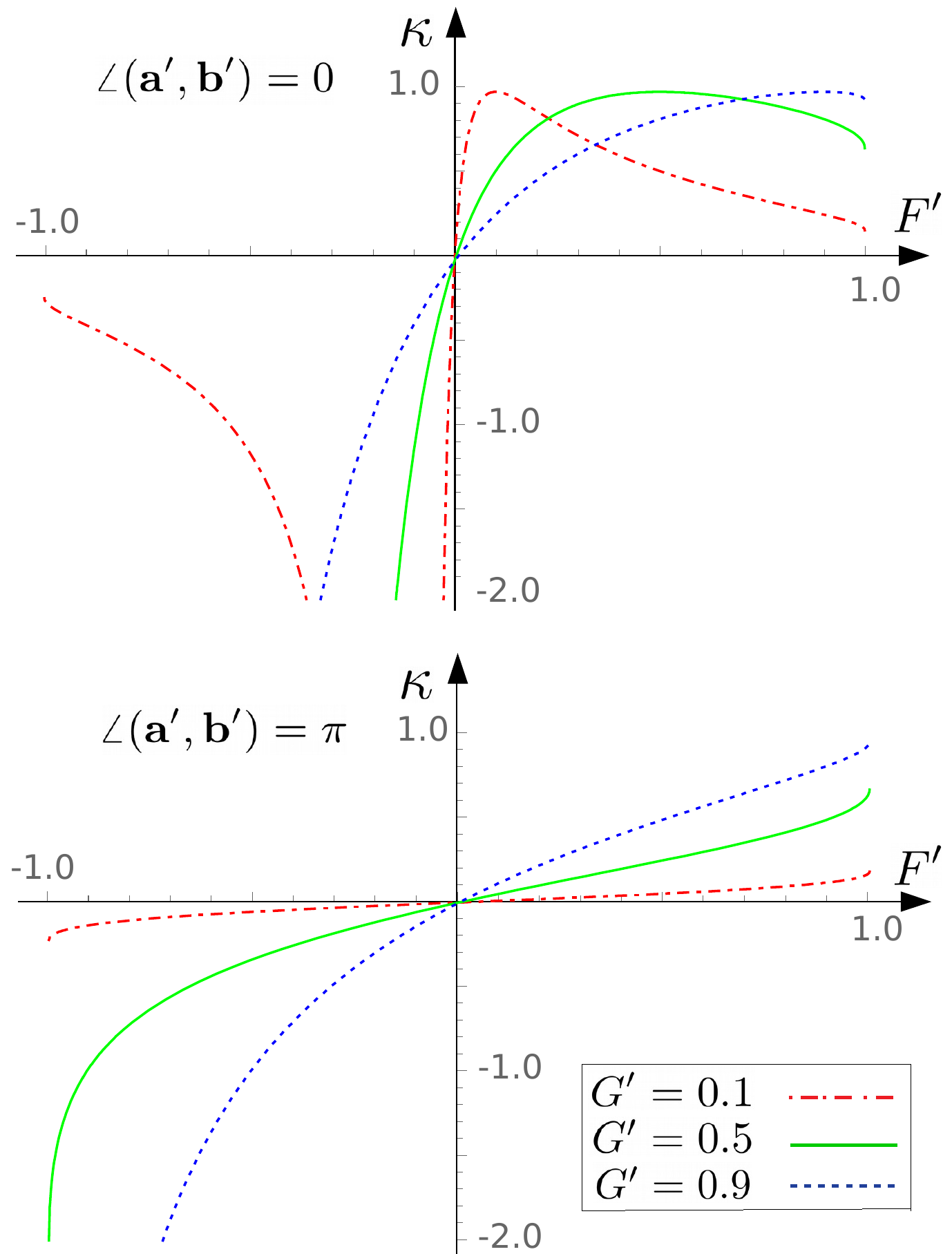}
\vspace{-0.1in} \caption{ The current $\kappa$ defined by (\ref{Current}) for different values of scalar moving modes $G^{\prime}$, $F^{\prime}$ and for angles $0$, $\pi$ between vectors $\textbf{a}^{\prime}$ and $\textbf{b}^{\prime}$.}
\label{fig:Current}
\end{figure}

Using relations (\ref{Normalization}), 
one can write down the current 
(\ref{TermsA}) as
\begin{equation}
\begin{gathered}
\label{Current}
\kappa = \frac{2 F^{\prime} G^{\prime} }{ 1 + F^{\prime} G^{\prime} - \textbf{a}^{\prime} \cdot \textbf{b}^{\prime}},
\end{gathered}
\end{equation}
which is shown in figure 
\ref{fig:Current} for different 
values of $F^{\prime}$ and 
$G^{\prime}$.

It is seen that if the 
left(or right)-moving mode of 
the current is independent of 
$\sigma_-$ (or $\sigma_+$), 
the expression (\ref{Current}) 
goes to zero
\begin{equation}
\label{ChiralLimit}
\kappa = 0, \quad \text{if}: \; \; F^{\prime}=0, \;  \; \; (\text{or} \; \; G^{\prime} = 0).
\end{equation}

The situation (\ref{ChiralLimit}) 
reproduces the chiral string 
properties, where only 
left(or right)-moving mode is 
allowed 
\cite{CarterPeter2, DavisKibblePicklesSteer, Blanco-PilladoOlumVilenkin, RybakAvgoustidisMartins2}.

\section{Junctions for transonic elastic strings}

To study Y-junctions for transonic elastic 
strings we start with the action for three 
connected current carrying strings 
\cite{SteerLilleyYamauchiHiramatsu} 
\begin{equation}
\begin{split}
\label{ActionJunctCurrent}
   S = - \sum_{i=1}^{3}  \mu_{ i} \int f(\kappa_i) \sqrt{-\gamma_i} \, \Theta \left( s_i(\tau) - \sigma_i \right)  d \sigma_i d \tau +\\
 + \sum _{i=1}^{3} \int \mathrm{f}_{\mu i} \left( x_i^{\mu} (s_i(\tau),\tau) - X^{\mu}(\tau) \right)  d \tau  + \\
 +  \sum_{i=1}^{3} \int \mathrm{g}_i \left( \varphi_i (s_i(\tau),\tau) - \Phi(\tau) \right) d \tau, 
\end{split}
\end{equation}
where the function $f(\kappa)$ is given 
by (\ref{F}), $\mu_{ i}$ are constants 
defined by the symmetry breaking scale, 
$\mathrm{f}_{\mu i}$, $\mathrm{g}_i$ 
are Lagrange multipliers for strings 
and currents, time functions  
$X^{\mu}(\tau)$ and 
$\Phi(\tau)$ define values for 
$x^{\mu}_i$ and $\varphi_i$ at the 
point where strings are connected, 
the index $i= 1,2,3$ denotes each 
of the three strings (the summation 
over index $i$ is carried out 
only when it is written explicitly).

Varying the action (\ref{ActionJunctCurrent}) 
with respect to $x_i^{\mu}$ and $\varphi_i$, 
we obtain the equations of motion 
(\ref{EqOfMotBPVOA}) and (\ref{EqOfMotBPVOB}) 
for each type of strings. Using (\ref{Det2}) 
and (\ref{EqOfMotBPVO4}) the boundary terms 
from equations of motion, which are 
proportional to $\delta(s_i(t) - \sigma_i)$, 
can be expressed as
\begin{equation}
\begin{split}
\label{BoundTerms}
& \quad \; \mu_i \eta^{ab} x_{i,a}^{\mu} \lambda_{b \, i} = \mathrm{f}_i^{\mu}, \\ 
&  2 \mu_i f_i f^{\prime}_{\kappa i} \eta^{ab} \varphi_{,a} \lambda_{b \, i} = \mathrm{g}_i,
\end{split}
\end{equation}
where $\lambda_{a \, i} = \left\lbrace \dot{s}_i,\; -1  \right\rbrace$.

The variation of the action (\ref{ActionJunctCurrent}) with respect 
to $\mathcal{X}_i^{\mu}$ and $\Phi$ 
gives us 
\begin{equation}
\begin{split}
\label{VarXPhi}   
& \sum_{i=1}^{3} \mathrm{f}_i^{\mu} = 0, \\ 
& \, \sum_{i=1}^{3} \mathrm{g}_i = 0,
\end{split}
\end{equation}
which can be rewritten using solutions 
(\ref{StrSolution}) and 
(\ref{CurrentSolution}) together with 
expressions (\ref{BoundTerms}) in the 
following way
\begin{equation}
\begin{split}
\label{Sum1} 
&  \sum_{i=1}^{3} \mu_i \left[ \textbf{a}_i^{\prime} (1+\dot{s}_i) - \textbf{b}_i^{\prime} (1-\dot{s}_i) \right] = 0, \\ 
& \sum_{i=1}^{3} \mu_i \left[ F_i^{\prime} (1+\dot{s}_i) -G_i^{\prime} (1-\dot{s}_i)  \right] = 0.
\end{split}
\end{equation}
Finally, variation of the action 
(\ref{ActionJunctCurrent}) with respect 
to $\mathrm{f}_i^{\mu}$ and $\mathrm{g}_i$ 
provides us the following relations
\begin{equation}
\begin{split}
\label{Varfg}
& x_i^{\mu}(s_i(\tau),\tau) = X^{\mu}(\tau), \\ 
& \; \varphi_i(s_i(\tau),\tau) = \Phi(\tau).
\end{split}
\end{equation} 
Differentiating (\ref{Varfg}), using 
the exact solutions (\ref{StrSolution}) 
and (\ref{CurrentSolution}) we obtain
\begin{equation}
\begin{split}
\label{VertCondD}
& \; (1+\dot{s}_i) \textbf{a}_{i}^{\prime} + (1-\dot{s}_i) \textbf{b}_i^{\prime} = 2\dot{\mathbf{X}}(t), \\ 
&  F_{ i}^{\prime} (1 + \dot{s}_i) + G_{ i}^{\prime}(1 - \dot{s}_i) = 2 \dot{\Phi}(t).
\end{split}
\end{equation}
Manipulating vectors 
$\textbf{a}^{\prime}_i$, 
$\textbf{b}^{\prime}_i$ and using 
(\ref{Sum1}) with (\ref{VertCondD}), 
it is possible to obtain the 
following equations

\begin{equation}
\begin{split}
\label{ABeq}  
&  \textbf{a}_{k}^{\prime} (1+\dot{s}_k) = \frac{2}{\mu} \sum_{i=1}^{3} (1-\dot{s}_i) \mu_i \textbf{b}_i^{\prime} - (1-\dot{s}_k) \textbf{b}_k^{\prime}, \\
& F_{ k}^{\prime} (1+\dot{s}_k) = \frac{2}{\mu} \sum_{i=1}^{3} (1-\dot{s}_i) \mu_i G_{ i}^{\prime} - (1-\dot{s}_k) G_{ k}^{\prime}
\end{split}
\end{equation}
and
\begin{equation}
\begin{split}
\label{Xeq}   
& \dot{\textbf{X}} = \frac{1}{\mu} \sum_{i=1}^{3} (1-\dot{s}_i) \mu_i \textbf{b}_i^{\prime}.
\end{split}
\end{equation}

Zero component of the vector equation in 
(\ref{BoundTerms}) provides energy 
conservation relation, which is 
identical to the standard Nambu-Goto 
scenario \cite{CopelandKibbleSteer}
\begin{equation}
\begin{split}
\label{EnergyConserv}  
& \sum_{i=1}^{3} \mu_i \dot{s}_i = 0.
\end{split}
\end{equation}
The relation (\ref{EnergyConserv}) does not 
provide an additional constraint, but 
is a consequence of 
equations of motion. Hence, the relation
(\ref{EnergyConserv}) can be used as
a check of numerical calculations that
are carried out below.

We parametrize the string worldsheets 
in such way that modes 
$\textbf{a}^{\prime}_i(\sigma_+)$, 
$F^{\prime}_i(\sigma_+)$ move outwards 
the string connection, while 
$\textbf{b}^{\prime}_i(\sigma_-)$ 
and $G^{\prime}_i(\sigma_-)$ move towards 
the string connection. Such choice means 
that $\textbf{b}^{\prime}_i(\sigma_-)$ 
and $G^{\prime}_i(\sigma_-)$ are initial 
values that define 
$\textbf{a}^{\prime}_i(\sigma_+)$ 
and $F^{\prime}_i(\sigma_+)$ by 
equations (\ref{ABeq}). 
The first three equations for 
vectors $\textbf{a}^{\prime}_i(\sigma_+)$ 
in (\ref{ABeq}) can be squared and using 
the normalization conditions 
(\ref{Normalization}) we eliminate 
$\textbf{a}_i^{\prime}(\sigma_+)$. 
Hence, we have the system of six 
independent algebraic equations 
(\ref{ABeq}) and six variables that 
can be found: three variables 
$\dot{s}_i$ and three variables 
$F_{i}^{\prime}(\sigma_+)$.

\begin{figure}[h]
\centering
		\includegraphics[scale=0.66]{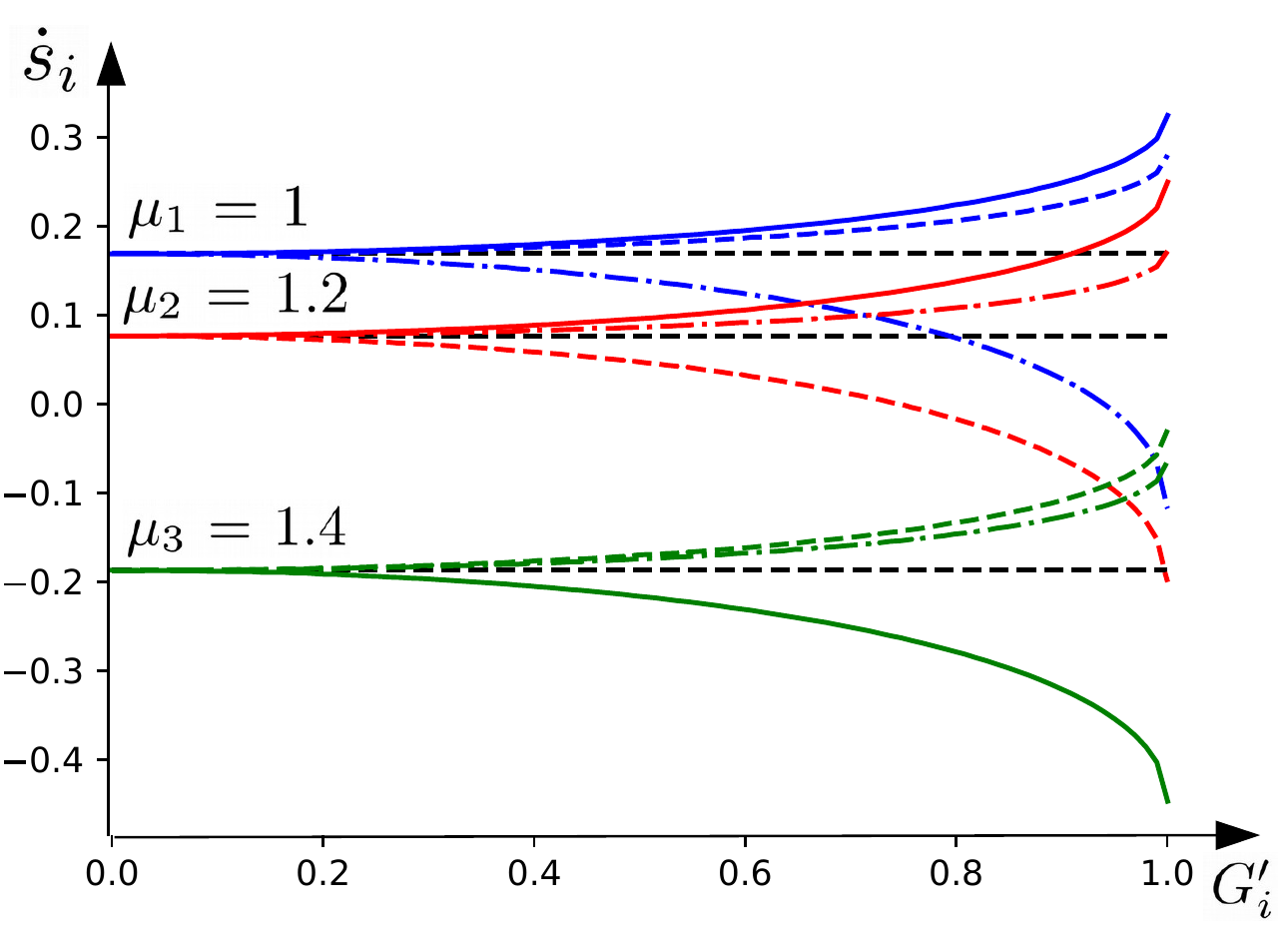}
\vspace{-0.1in} \caption{ Dynamics of Y-junctions represented by $\dot{s}_i$ of strings with $\mu_1=1$ (blue), $\mu_2=1.2$ (red), $\mu_3=1.4$ (green) and oriented with angles $2\pi/3$ between them. Dashed lines show the values of $\dot{s}_i$ depending on $G^{\prime}_1$, dash-dotted on $G^{\prime}_2$, solid on $G^{\prime}_3$. Black dashed lines demonstrate no changes of $\dot{s}_i$ when all $G_i^{\prime}$ increase simultaneously. }
\label{fig:JuncDynamics}
\end{figure}

It is illustrative to compare values 
of $\dot{s}_i$ for strings with currents 
and without. For this purpose we fix 
angles between 
$\textbf{b}^{\prime}_i(\sigma_-)$, 
define string constants $\mu_i$ and 
evaluate the system of equations 
(\ref{ABeq}) for different values of 
$G_i^{\prime}(\sigma_-)$. 
An example of such dependence is shown 
in figure \ref{fig:JuncDynamics}.

The description above demonstrates 
that dynamics of Y-junctions for 
transonic elastic strings can be 
described within Nambu-Goto approximation.

\subsection{Kinks for elastic strings} \label{Kinks for elastic strings}

Having considered the Y-junctions,
we can treat the formation and evolution 
of kinks for elastic strings. To do so
we need simply change the sum in 
previous equations for $2$ 
strings instead of $3$.
Let us consider the situation
when parameters for strings are 
$\mu_1=\mu_2$. Hence, from equations 
(\ref{VertCondD}) and (\ref{ABeq}) 
one can deduces the following relations 
for two possible situations that
satisfy conditions for the kink
\begin{subequations}
\begin{align}
   \label{KinksRelationsA}
& \dot{s}_1 = -1 = - \dot{s}_2, \; F^{\prime}_2 = G^{\prime}_1,\\ 
     \label{KinksRelationsB}
    \text{or } & \; \dot{s}_1 = 1 = - \dot{s}_2, \; F^{\prime}_1 = G^{\prime}_2.
\end{align}
\end{subequations}

\begin{figure}[h]
\centering
		\includegraphics[scale=0.75]{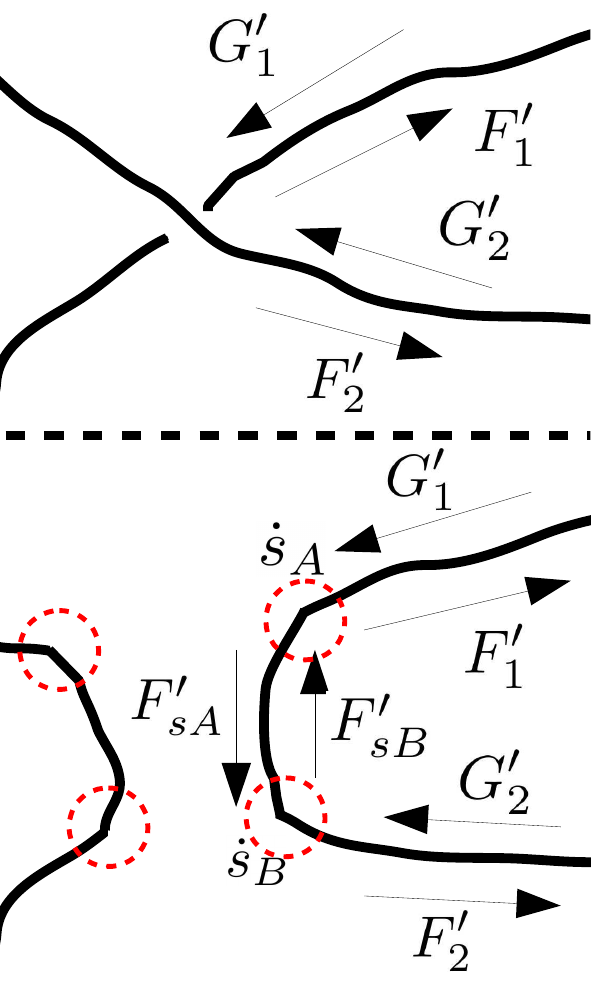}
\vspace{-0.1in} \caption{ Upper panel demonstrates two cosmic strings before the collision. Arrows with $G^{\prime}$ and $F^{\prime}$ define left-right-moving modes of currents on cosmic strings. Bottom panel represents the situation when collided strings intercommute, exchange their moving modes.}
\label{fig:Kinks}
\end{figure}

Illustrative example of two strings 
intercommutation, is shown on figure 
\ref{fig:Kinks}. 
After collision two kinks 
propagating in opposite direction 
are formed on each of strings. 
The kink $\dot{s}_A$ corresponds
to situation (\ref{KinksRelationsA})
with $F^{\prime}_{sA} = G_1^{\prime}$,
while another kink $\dot{s}_B$ to 
(\ref{KinksRelationsB})
with $F^{\prime}_{sB} = G_2^{\prime}$.

The velocities of kinks
follow from equation (\ref{Xeq}), that is
\begin{subequations}
\begin{align}
   \label{KinksVelA}
& \dot{\textbf{X}}^2_A = (1-G^{\prime \, 2}_1),\\ 
     \label{KinksVelB}
    & \dot{\textbf{X}}^2_B = (1-G^{\prime \, 2}_2),
\end{align}
\end{subequations}
which are different from the speed of light
if the correspondent current component 
is non-zero.

It should be noted that if an elastic 
string collides with a standard
Nambu-Goto (or chiral 
with $G_{1}^{\prime}=0$) string, 
the intermediate growing section, 
between kinks $\dot{s}_A$ 
and $\dot{s}_B$, is
described by the chiral model.

In the same manner 
kinks appear when colliding 
strings form Y-junction:
from the discontinuity of 
corresponding modes. 
The formation of kinks for 
elastic strings qualitatively is
identical to the standard Nambu-Goto
model considered in 
\cite{CopelandKibbleSteer2}, except
of the fact that the speed of kinks
propagation is not equal to the 
speed of light, but given by
(\ref{KinksVelA}) and (\ref{KinksVelB}).

\section{Collisions of transonic elastic strings} \label{Collisions of transonic elastic strings}

It is always possible to chose small region, 
where collided strings can be considered 
straight. We are going to study kinematic 
conditions for straight strings to produce 
a Y-junction.

We decompose the straight string solution 
as a linear combination of "bare" and 
current carrying parts 
\cite{RybakAvgoustidisMartins2} 
\begin{equation}
\begin{split}
\label{StrStringCurrent0}
 & \textbf{x}_{i} = \textbf{y}_{i} + \textbf{z}_{i}, 
\end{split}  
\end{equation}
where the ``bare'' part is given by
\begin{equation}
\begin{split}
\label{StrStringCurrent1}
 & \textbf{y}_{1,2} = \left\lbrace -\gamma^{-1}_v \sigma \cos \alpha; \,\mp \gamma^{-1}_v \sigma \sin \alpha; \, \pm  \upsilon \tau \right\rbrace, \\
 & \quad \; \textbf{y}_{3} = \left\lbrace \gamma^{-1}_u \sigma \cos \theta; \, \gamma^{-1}_u \sigma \sin \theta; \, u \tau \right\rbrace,
\end{split}  
\end{equation}
while the current carrying part is 
described by
\begin{equation}
\begin{split}
\label{StrStringCurrent2}
 & \textbf{z}_{i} = -g_{i} \sigma_- (\dot{\textbf{y}}_{i} - \textbf{y}_{i}^{\prime} ) -f_{i}\sigma_+ (\dot{\textbf{y}}_{i} + \textbf{y}_{i}^{\prime} ),
\end{split}  
\end{equation}
with $\gamma^{-1}_v = \sqrt{1-v^2}$.

Constants $f_{i}$ and $g_{i}$ in 
(\ref{StrStringCurrent2}) represent 
the current contribution for 
left- and right-moving modes.

From (\ref{StrStringCurrent0})-(\ref{StrStringCurrent2}) one can 
find that 
\begin{equation}
\begin{split}
\label{ABStraight}
& \textbf{a}^{\prime}_{i} = (1-2 f_i) (\dot{\textbf{y}}_i + \textbf{y}^{\prime}_i), \quad |\textbf{a}^{\prime}_i|^2 = (1 - 2 f_i)^2, \\
& \textbf{b}^{\prime}_{i} = (1-2 g_i) (\dot{\textbf{y}}_i-\textbf{y}^{\prime}_i),  \quad |\textbf{b}^{\prime}_i|^2 = (1 - 2 g_i)^2.
\end{split}
\end{equation}

Comparing constants $f_i$ and $g_i$ in 
(\ref{ABStraight}) with 
(\ref{Normalization}), we establish 
the relations
\begin{equation}
\begin{gathered}
\label{RelationFFfg}
f_i = \frac{1}{2} \left( 1 - \sqrt{1-F_{ i}^{\prime \, 2}} \right),  \\ g_i = \frac{1}{2} \left( 1 - \sqrt{1-G_{ i}^{\prime \, 2}} \right).
\end{gathered}
\end{equation}

Parameters $v$ and $\alpha$ define 
orthogonal velocity and orientation 
of non-current carrying
string \cite{CopelandKibbleSteer2},
but for elastic (and chiral 
\cite{Blanco-PilladoOlumVilenkin}) 
strings 
it is not the case. It happens due 
to presence of longitudinal component 
of velocity in considered 
parametrization. Hence, we will treat
$v$ and $\alpha$ as some constant 
parameters combination of which
provide string velocity and orientation.
Alteration of
parameters interpretation  
does not affect the
validity of applied method.

In order to find out for which values of 
$v$ and $\alpha$, the third string 
can be produced (which means that 
$\dot{s}_3>0$), we need to derive the 
orientation and velocity of 
a newly created string, i.e. 
$\theta$ and $u$ parameters. 
To obtain these variables, 
we follow the procedure of \cite{CopelandKibbleSteer2}, i.e. we 
write down the expression for 
$\dot{\textbf{X}}$, given in 
(\ref{Xeq}), by substituting 
$\sigma \rightarrow s_3(\tau)$ 
in
(\ref{StrStringCurrent1}) and 
(\ref{StrStringCurrent2})
\begin{equation}
\label{XdotStraight}
 \dot{\textbf{X}} = \left\lbrace T_1(\tau) \gamma_u^{-1} \cos \theta; \; T_1(\tau) \gamma_u^{-1} \sin \theta; \; T_2(\tau) u  \right\rbrace,
\end{equation}
where $T_1(\tau) = \dot{s}_3(\tau) + g_3 (1-\dot{s}_3(\tau)) - f_3(1+\dot{s}_3(\tau))$ 
and 
$T_2(\tau) = 1-f_3(1+\dot{s}_3(\tau))+g_3(1-\dot{s}_3(\tau))$.

Combining (\ref{XdotStraight}) with 
(\ref{Xeq}) one can obtain the vector 
equation, from which $\theta$ and 
$u$ are determined via 
$\textbf{b}^{\prime}_i$. 

\begin{figure}[h]
\centering
		\includegraphics[scale=0.62]{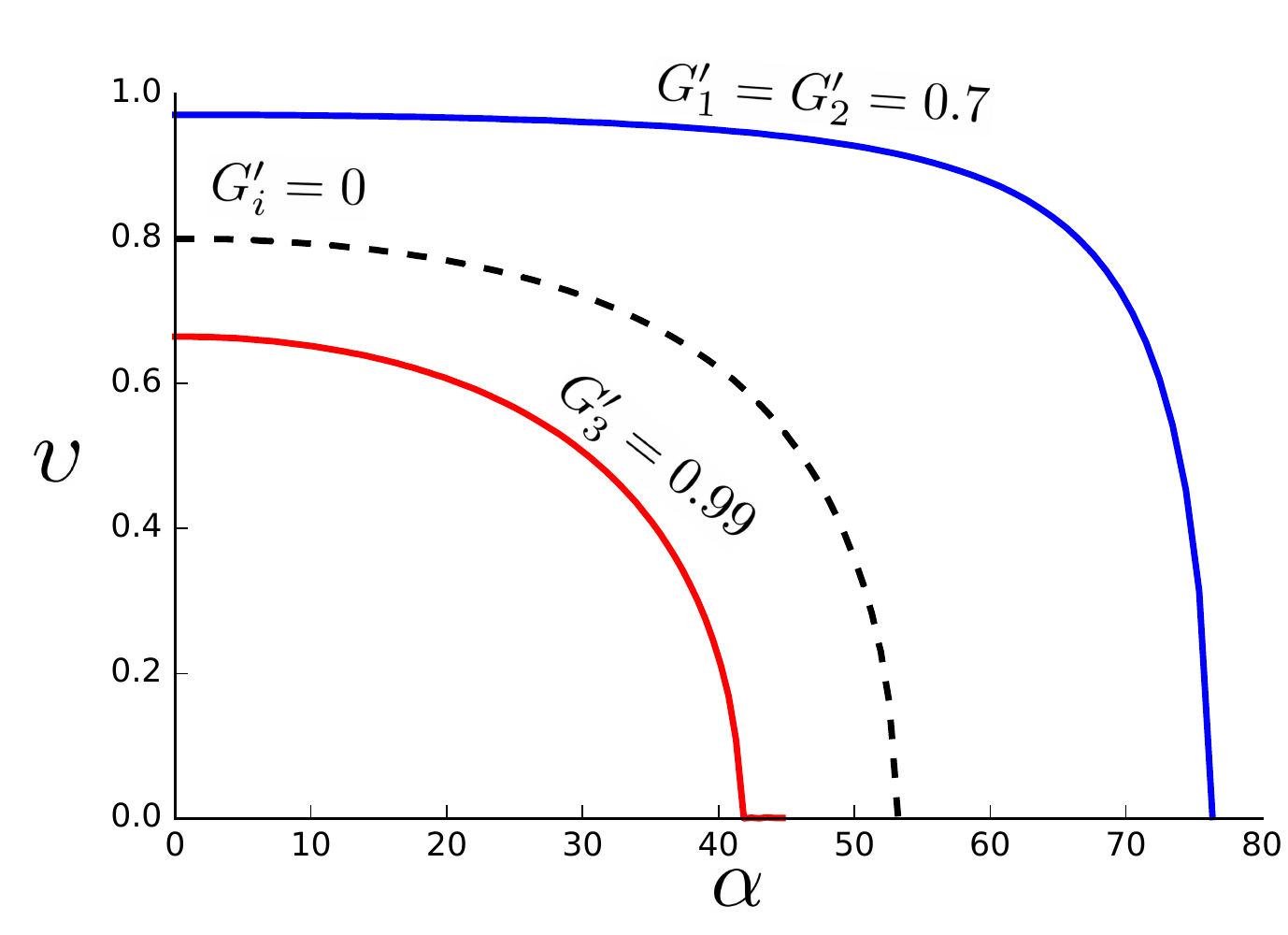}
\vspace{-0.1in} \caption{Symmetric case. Range of parameters $v$ and $\alpha$, which allow for colliding strings with $\mu_1=\mu_2=1$ to produce the Y-junction ($\dot{s}_3>0$ corresponds to areas below lines) with $\mu_3=1.2$. The solid blue line corresponds to $G^{\prime}_1=G^{\prime}_2=0.7$, red line to $G^{\prime}_3=0.99$, while all others $G_i^{\prime}$ are zeros. Dashed black line represents the case when all $G^{\prime}_i=0$. }
\label{fig:StringCollisionEq}
\end{figure}

\begin{figure}[h]
\centering
		\includegraphics[scale=0.68]{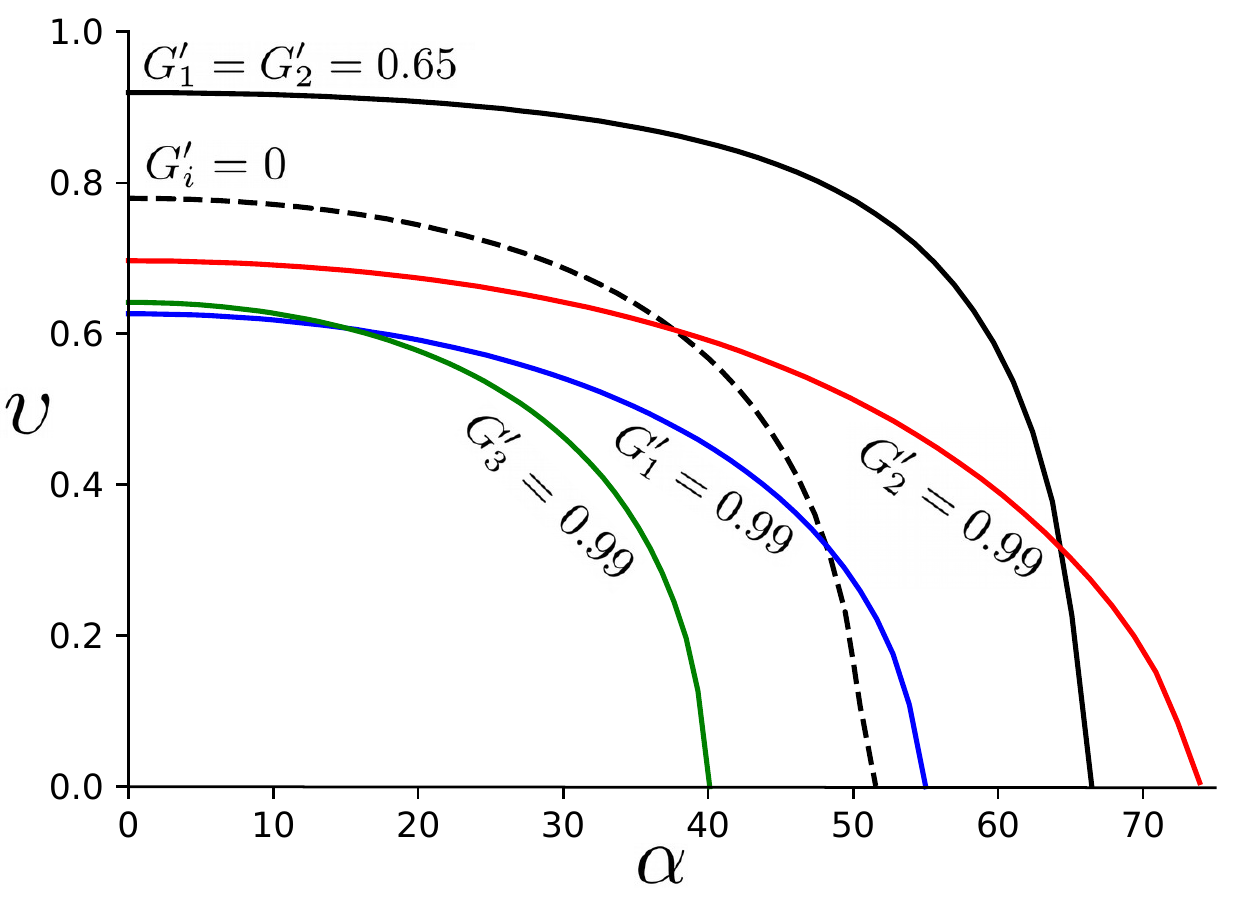}
\vspace{-0.1in} \caption{ Asymmetric case. Range of parameters $v$ and $\alpha$, which allow for colliding strings with $\mu_1=1$ and $\mu_2=1.2$ to produce the Y-junction ($\dot{s}_3>0$ corresponds to areas below lines) with $\mu_3=1.4$. The solid blue line corresponds to $G^{\prime}_1=0.99$, red line to $G^{\prime}_2=0.99$, green line to $G^{\prime}_3=0.99$ while all others $G_i^{\prime}$ are zeros. Dashed black line represents the case when all $G^{\prime}_i=0$, while black solid when $G^{\prime}_1=G^{\prime}_2=0.65$, $G^{\prime}_3=0$. }
\label{fig:StringCollision}
\end{figure}

To summarize, we have nine equations: 
six equations from (\ref{ABeq}) and 
three equations from (\ref{Xeq}). 
Therefore, we can derive eight 
variables $F^{\prime}_i$, 
$\dot{s}_i$, $u$, $\theta$ 
defining another eight variables 
$\mu_i$, $G_i^{\prime}$, $v$, $\alpha$. 
The vector equality (\ref{Xeq}) 
with (\ref{XdotStraight}) does not 
provide three independent equations, 
but only two, similarly as in 
\cite{CopelandKibbleSteer2}. 
Having all this information, we 
can numerically solve this system 
of algebraic equations. As a 
result, one can obtain the region 
of values $v$ and 
$\alpha$ for which colliding 
strings give rise to Y-junctions 
($\dot{s}_3>0$), 
see figure 
\ref{fig:StringCollisionEq}
for symmetric string
collision and 
figure 
\ref{fig:StringCollision}
for asymmetric string collision.

Production of Y-junction also leads 
to creation of kinks 
\begin{equation}
\label{StrStringCurrentKinks}
  \textbf{K}_{1,2} = (1 - 2 g_{1,2}) \left\lbrace \gamma^{-1}_v \cos \alpha; \,\pm \gamma^{-1}_v  \sin \alpha; \, \pm  \upsilon  \right\rbrace \tau
\end{equation}
that propagate along collided
strings, similarly as it 
happens for standard
non-current Nambu-Goto strings
\cite{CopelandKibbleSteer2}.

It is important to note that
for all strings 
the constants
$\mu_i$ were fixed, and we assumed
that there is a relation between
tensions of 
strings $T_i$.
This assumption
allows us to eliminate one 
degree of freedom and treat
$G^{\prime}_3$ as known 
value (it might be done through
relation (\ref{Current})).
Possible bound between tensions
of current carrying strings
needs further investigation for
particular models and goes 
beyond the scope of this 
paper.

\section{Conclusions} \label{Conclusions}

We revisited the exact solution for
elastic transonic strings in 
Minkowski space 
\cite{Carter90, Carter95} 
with the method developed in 
\cite{Blanco-PilladoOlumVilenkin}. 
The exact solution allowed us 
to consider left- and right-moving 
modes, which made it possible 
to treat the dynamics of 
Y-junctions in a similar 
manner as it was done in 
\cite{CopelandKibbleSteer2}.

The system of equations 
(\ref{ABeq}) allowed us to 
obtain the rate of 
string lengths change $\dot{s}_i$,
see figure \ref{fig:JuncDynamics}, 
requiring the definition of 
incoming components of the 
current $G^{\prime}_i$. 
The values of incoming 
current components $G^{\prime}_i$ 
should be determined by strings 
properties. Thus, in the case 
of cosmic superstrings the 
values of $G^{\prime}_i$ 
might be defined similarly 
to saturated BPS state 
(see \cite{PolchinskiCopelandMyers, DasguptaMukhi} 
for details), given by
$$\mu_{p,q} = \mu_F \sqrt{(p-q C_0)^2 + q^2/g_s^2}.$$
For superconducting and wiggly 
cosmic strings with Y-junctions, 
the definitions of $G^{\prime}_i$ 
should arise from the values of 
tensions and mass per unit lengths (\ref{UTcurrent}). 
The exact definition of $G^{\prime}_i$ 
for particular type of strings 
needs further investigation and 
goes beyond the scope of this 
paper.

We studied the kink dynamics
for elastic strings in section
\ref{Kinks for elastic strings}. 
We obtained values of $\dot{s}_i$ 
that are essential for existence
of kink-like discontinuity.
We also demonstrated an 
example of elastic strings 
intercommutation
and determined the velocities 
of these kinks.

In section 
\ref{Collisions of transonic elastic strings} 
we found kinematic constraints that 
should be satisfied to give rise 
to a Y-junction for elastic strings. 
In particular, we 
obtained a range of parameters $v$,
$\alpha$ and $G_i^{\prime}$ of 
collided strings
(\ref{StrStringCurrent0}) 
for which $\dot{s}_3>0$. The symmetric 
case of elastic strings collision 
is shown in figure 
\ref{fig:StringCollisionEq} and 
asymmetric case is shown
in figure
\ref{fig:StringCollision}.

\begin{acknowledgments}

This work was supported by FCT - Fundação para a Ciência e a Tecnologia through national funds (PTDC/FIS-PAR/31938/2017) and by FEDER - Fundo Europeu de Desenvolvimento Regional through COMPETE2020 - Programa Operacional Competitividade e Internacionalização (POCI-01-0145-FEDER-031938). We also would like to thank Juliane F. Oliveira, Lara G. Sousa, Ricardo C. Costa, Carlos J.A.P. Martins and Anastasios Avgoustidis for fruitful discussions and help in organizing the manuscript. 

\end{acknowledgments}

\section*{Appendix: Comparison with \cite{SteerLilleyYamauchiHiramatsu}} \label{Appendix}

In section 
\ref{Collisions of transonic elastic strings} 
we considered the formation of Y-junctions 
for transonic model of straight strings 
determined by the action (\ref{Action})
with $f(\kappa)$ defined in (\ref{F}).
It was shown that for collision of 
transonic
elastic straight strings there are
kinematic conditions that should be
satisfied to form a Y-junction.
In our study we did
not face overdetermined system
of equations, in contrast to 
result in
\cite{SteerLilleyYamauchiHiramatsu}.
The general approach of 
\cite{SteerLilleyYamauchiHiramatsu}
claims that for 
\textit{magnetic} 
and \textit{electric}  
superconducting strings the formation 
of Y-junction is impossible.
On the other hand, the result in
section 
\ref{Collisions of transonic elastic strings}
states that formation of Y-junction
is possible 
for a particular type of current, 
described by transonic elastic model.
This appendix is intended to provide
detailed comparison of
our result with result obtained in
\cite{SteerLilleyYamauchiHiramatsu}.
For the sake of clarity we
denote equations related to work 
\cite{SteerLilleyYamauchiHiramatsu}
as (...$^*$).

\subsection{Comparison of equations} \label{Comparison of equations}

Let us write down equations of motion for 
Y-junction used in work
\cite{SteerLilleyYamauchiHiramatsu}
(above equations (16$^*$)-(17$^*$) in
\cite{SteerLilleyYamauchiHiramatsu})
\begin{subequations}
\label{Ap0}
\begin{align}
   \label{Eq160}
    &  \partial_a \left( \sqrt{-\gamma_i} T_i^{a b} x^{\mu}_{i , b} \Theta \left( \tilde{s}_i(\tau_i) -\tilde{\sigma}_i  \right) \right) = \nonumber \\
& \qquad \qquad \qquad \qquad \qquad \qquad  = \mathrm{f}_i^{\mu} \delta(\tilde{s}_i(\tau_i) - \tilde{\sigma}_i),\\ 
     \label{Eq170}
    & \quad \partial_a \left(  \sqrt{-\gamma_i} z^a_i \Theta \left( \tilde{s}_i(\tau_i) -\tilde{\sigma}_i  \right)  \right) = \mathrm{g}_i \delta(\tilde{s}_i(\tau_i) - \tilde{\sigma}_i),
\end{align}
\end{subequations}
where  
$z_i^a = \sqrt{\kappa_{0 i}} c^a_i $,
$c_i^a = - 2  f^{\prime}_{\kappa_i} \gamma_i^{a b} \varphi_{i,b} $, 
$ \sqrt{-\gamma_i} T^{a b}_i = \mathcal{T}^{a b}_i $, $\kappa_{0 \, i}$ is a constant 
multiplier and all derivatives are taken with 
respect to conformal gauge parameters $\tau_i$
and $\tilde{\sigma}_i$, 
which are different 
for each string (in contrast 
to the gauge of the present study, where 
$\tau$ is the same for all strings) and
provide the relations
$$ \partial_{\tau_i} x_i^{\mu} \partial_{\tilde{\sigma}_i} x_{i \, \mu} = 0, \quad (\partial_{\tau_i} x_i^{\mu})^2 = - (\partial_{\tilde{\sigma}_i} x_{i}^{\mu})^2.$$
Equations (\ref{Ap0}) are parametrization 
invariant, namely one can chose any
$\tilde{\sigma}_i$ and $\tau_i$.
If one choses the gauge ($\tau$, $\sigma_i$)
of this work and uses expressions
(\ref{Transform}), (\ref{EqOfMotBPVO3})
for elastic strings, defined
by function (\ref{F}),
the system of equations 
(\ref{Ap0}) is reduced to
\begin{subequations}
\label{Ap0My2}
\begin{align}
   \label{Eq16My2}
    &   \partial_a \left( \eta^{ab} x^{\mu}_{i , b} \Theta \left( s_i(\tau) -\sigma_i  \right) \right) = \mathrm{f}_i^{\mu} \delta(s_i - \sigma_i),\\ 
     \label{Eq17My2}
    & \, \partial_a \left(  \eta^{ab} \varphi_{i , b} \Theta \left( s_i(\tau) -\sigma_i  \right)  \right) = \mathrm{g}_i \delta( s_i - \sigma_i).
\end{align}
\end{subequations}
Substituting exact solutions
(\ref{StrSolution}), (\ref{CurrentSolution})
in (\ref{Ap0My2}) and using condition 
(\ref{VarXPhi})
one can see that boundary 
terms of equations (\ref{Ap0My2}) 
are identical to equations (\ref{Sum1}).

In case of conformal
gauge ($\tilde{\sigma}_i, \tau_i$), 
boundary terms of equations 
(\ref{Ap0})
have the form of equations 
(16$^*$)-(17$^*$)
of reference 
\cite{SteerLilleyYamauchiHiramatsu},
given by
\begin{subequations}
\label{Ap1}
\begin{align}
   \label{Eq16}
    &  \sum_i \sqrt{-\gamma_i}  \left( T_i^{0 b}  \dot{\tilde{s}}_i - T_i^{1 b}  \right) x^{\mu}_{i , b} = 0,\\ 
     \label{Eq17}
    & \quad \quad \sum_i  \sqrt{-\gamma_i}  \left( z^0_i \dot{\tilde{s}}_i - z^1_i \right) = 0,
\end{align}
\end{subequations}
where $\dot{\tilde{s}}_i \equiv \frac{d \tilde{s}_i}{d \tau_i} $.

We demonstrated that equations (16$^*$)-(17$^*$) 
for Y-junction in  
\cite{SteerLilleyYamauchiHiramatsu}
coincide with equations (\ref{Sum1})
of the main text. There is a full
agreement between equations of 
\cite{SteerLilleyYamauchiHiramatsu}
and equations of the manuscript,
see table \ref{EqsComparison}
for correspondence.
Equations (25$^*$), (28$^*$) and (29$^*$) 
of \cite{SteerLilleyYamauchiHiramatsu},
should be identical to (\ref{Ap1}) and 
to (\ref{Sum1}), but just
written in a preferred rest-frame.
Hence, the amount of equations 
in our study and 
work \cite{SteerLilleyYamauchiHiramatsu}
are the same.

\begin{center}
\begin{table} 
\begin{tabular}{ |r|l| } 
 \hline  
 Gauge of this work & Conformal gauge in \cite{SteerLilleyYamauchiHiramatsu} \\ 
 \hline
 Eq.(\ref{VarXPhi}) =&= Eq.(15$^*$)  \\ 
 Eq.(\ref{Sum1}) =&= Eq.(14$^*$)  \\ 
 Eq.(\ref{Varfg}) =&= Eqs.(16$^*$)-(17$^*$)  \\
 \hline 
\end{tabular}
\caption{\label{EqsComparison} Correspondence between equations of the manuscript and work \cite{SteerLilleyYamauchiHiramatsu}}
\end{table}
\end{center}

\subsection{Straight string solution and number of unknown variables}
\label{Straight string solution and number of unknown variables}

To understand where the disagreement
with \cite{SteerLilleyYamauchiHiramatsu}
comes from, we also need to count the number
of unknown variables, since the number
of equations in our study and
in work 
\cite{SteerLilleyYamauchiHiramatsu}
are the same.

To start, let us write down the solution
for straight strings, which satisfies
equations of motion
(\ref{EqOfMot11}) and can
be considered as a linear term of
Taylor expansion close to the point 
of strings collision, i.e. for 
gauge of this work
\begin{equation}
\begin{gathered}
\label{StraightStringSolution}
x^{\mu}_i(\sigma_i,\tau) = A_i^{\mu} \sigma_i + B_i^{\mu} \tau + \mathcal{O}(\sigma_i^2, \tau^2), \\
\varphi_i(\sigma_i,\tau) = C_i \sigma_i + D_i \tau + \mathcal{O}(\sigma_i^2, \tau^2) \\
\text{or for conformal gauge} \\
x^{\mu}_i(\tilde{\sigma}_i,\tau_i) = \tilde{A}^{\mu} \tilde{\sigma}_i + \tilde{B}^{\mu} \tau_i + \mathcal{O}(\tilde{\sigma}_i^2, \tau_i^2), \\
\varphi_i(\tilde{\sigma}_i,\tau_i) = \tilde{C}_i \sigma_i + \tilde{D}_i \tau_i + \mathcal{O}(\tilde{\sigma}_i^2, \tau_i^2).
\end{gathered}
\end{equation}
where $A_i^{\mu}$, $B_i^{\mu}$, 
$\tilde{A}_i^{\mu}$, $\tilde{B}_i^{\mu}$,
$C_i$, $D_i$, $\tilde{C}_i$ and 
$\tilde{D}_i$ are constants 
with possibly different physical
meaning. The form of solution
(\ref{StraightStringSolution}) 
provides left- and right-moving modes, 
similarly to (\ref{StrSolution}) and
(\ref{CurrentSolution}). 

Mass per unit length $U$ and tension
$T$, given by (\ref{UTcurrent}),
are dynamical parameters that are
constructed from left- and
right-moving modes of 
(\ref{StraightStringSolution}) for the
corresponding gauge. When 
$U$, $T$ (or current $\kappa$ with action) 
are fixed 
one can still chose
different left- and right-moving 
modes. 
This fact is well seen from 
expression (\ref{Current}), where
the same value of the current $\kappa$
can be constructed by different 
values of $F^{\prime}$ and 
$G^{\prime}$. 

\begin{figure}[h]
\centering	\includegraphics[scale=0.95]{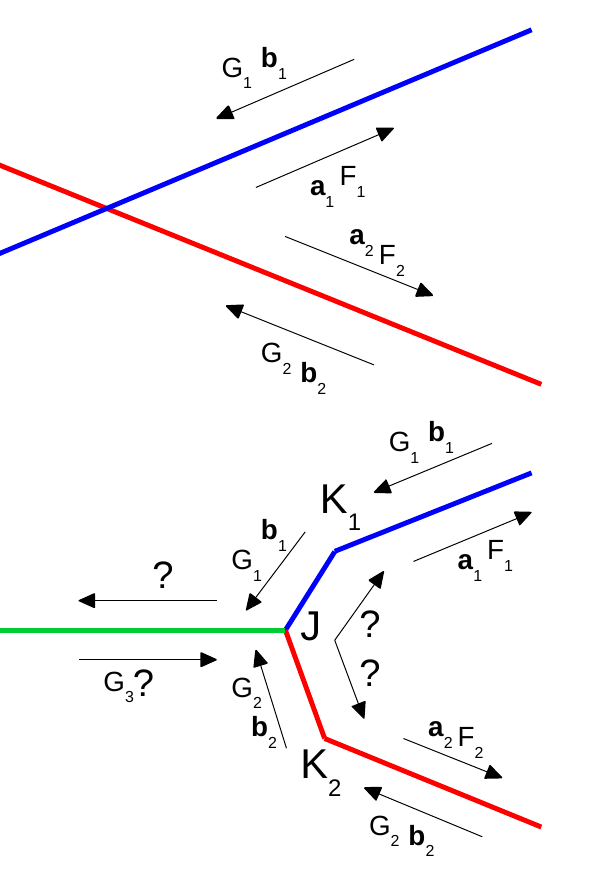}
\vspace{-0.1in} \caption{ Schematic picture of 
left- and right-moving modes of elastic 
transonic strings for the gauge 
described in the main text, section
\ref{Solution in Minkowski space for transonic elastic strings}. 
Upper panel shows left- and
right-moving modes before the collision
of strings, lower panel shows modes after.  
Moving modes with question symbol should be 
determined by equations for Y-junction 
(\ref{ABeq})-(\ref{Xeq}).}
\label{fig:StringCollisionSchem}
\end{figure}

Expressing all parameters, such as 
mass per unit length $U$, tension $T$ and 
current $\kappa$
by left- and right-moving modes,
we can count number of unknown 
variables.
All derivatives of outgoing modes, 
can be determined
through conditions (\ref{Varfg}),
introducing $\dot{\Phi}$ and
$\dot{\textbf{X}}$. The last 
variable can be substituted by
expression with 
parameters $\alpha$, $u$ and $\dot{s}_3$ of
newly created string. Hence,
there are $4$-equations from
(\ref{Ap1}) and energy conservation
condition (which is automatically
satisfied), $2$-independent
equations come from first 
expression of (\ref{Varfg}),
providing $6$ independent equations 
for six variables: $u$, $\alpha$, 
$\dot{s}_i$ and $\dot{\Phi}$. 
Thus, expressing all parameters as 
left- and right-moving
modes one obtains the system of equations,
which has the same amount of equations
and unknown variables 
(as it is mentioned in the end of section 
\ref{Collisions of transonic elastic strings}
one also needs to define incoming mode $G_3$).

\subsection{The difference with work \cite{SteerLilleyYamauchiHiramatsu}}

The process of strings collision
for elastic transonic strings is 
illustrated in figure 
\ref{fig:StringCollisionSchem}. The upper
panel demonstrates left-, right-moving
modes of strings before
the collision, the lower - after.
Modes denoted by question symbol 
should be obtained
from Y-junction equations at $J$.
Two kinks, illustrated in figure
\ref{fig:StringCollisionSchem} 
as $K_1$ and $K_2$, are formed
after strings collision and, according
to section \ref{Kinks for elastic strings},
incoming modes propagate without modification
through kinks $K_1$ and $K_2$ for
special choice of parametrization, 
i.e. modes
$G_1$, $\textbf{b}_1$, $G_2$ and 
$\textbf{b}_2$ can pass directly to 
Y-junction. Kinks for elastic transonic 
straight strings
(\ref{StrStringCurrent0})
are described by (\ref{StrStringCurrentKinks}).
Schematic figure
\ref{fig:StringCollisionSchem}
of strings collision 
with particular choice of parametrization
is valid
for elastic transonic strings as well
as for standard Nambu-Goto strings 
\cite{CopelandKibbleSteer,CopelandKibbleSteer2},
and chiral strings
\cite{RybakAvgoustidisMartins2}.

For conformal gauge choice, which was
used in \cite{SteerLilleyYamauchiHiramatsu},
incoming modes for $J-K_1$ and $J-K_2$, 
in general, are not the same as
modes before kinks and should be determined 
by equations
for kink discontinuity. As a result, 
to obtain
similar parameter region space,
which allows Y-junction formation,
one should solve the system of equations 
that includes equations for kinks
$K_1$, $K_2$ and junction $J$ 
simultaneously. 
This fact demonstrates that 
the gauge choice, which was used in this
work, simplifies equations 
allowing to consider incoming modes 
for Y-junction the same as modes before
$K_1$ and $K_2$ kinks.   

To summarize the comparison, it was
demonstrated in section 
\ref{Comparison of equations}
that equations 
for Y-junctions of this study and work
\cite{SteerLilleyYamauchiHiramatsu}
are in agreement,
one system of
equations can be transformed
to the other, see table \ref{EqsComparison}.
In section 
\ref{Straight string solution and number of unknown variables}
it was shown that any type of straight strings 
can be represented by the form 
(\ref{StraightStringSolution}) and 
split for left- and right-moving modes.
Defining string parameters,
such as $T$, $U$ and current through 
left- and right-moving modes, one obtains
the system of equations that has 
the same amount of unknown variables
and equations for 
Y-junction.
Hence, equations of work
\cite{SteerLilleyYamauchiHiramatsu}
written via left- and right-moving modes
(which do not require 
separated consideration
of \textit{electric} and 
\textit{magnetic} types of current) 
are reduced to (\ref{ABeq}), (\ref{Xeq})
providing the same range of 
parameters that allows formation
of Y-junction. 
In study 
\cite{SteerLilleyYamauchiHiramatsu}
equations are not written in terms of
left- and right-moving modes, but in 
terms of tension $T$, mass per unit length
$U$ and \textit{electric} 
(or \textit{magnetic}) current. 
Each value of $U$ and $T$ (as well as 
current) can
be represented by different 
left- and right-moving modes.
It means that there are more unknown
variables in equations written via
left- and right-moving modes than in 
equations written via $U$ and $T$.
Fixing mass per unit
length $U$ and tension $T$, as it was done
in \cite{SteerLilleyYamauchiHiramatsu},
one also fixes corresponding outgoing and 
incoming modes 
obtaining overdetermined system 
of equations,
i.e. less unknown variables
than equations for Y-junction.

\bibliography{biblio}

\begin{thebibliography}{56}%
\makeatletter
\providecommand \@ifxundefined [1]{%
 \@ifx{#1\undefined}
}%
\providecommand \@ifnum [1]{%
 \ifnum #1\expandafter \@firstoftwo
 \else \expandafter \@secondoftwo
 \fi
}%
\providecommand \@ifx [1]{%
 \ifx #1\expandafter \@firstoftwo
 \else \expandafter \@secondoftwo
 \fi
}%
\providecommand \natexlab [1]{#1}%
\providecommand \enquote  [1]{``#1''}%
\providecommand \bibnamefont  [1]{#1}%
\providecommand \bibfnamefont [1]{#1}%
\providecommand \citenamefont [1]{#1}%
\providecommand \href@noop [0]{\@secondoftwo}%
\providecommand \href [0]{\begingroup \@sanitize@url \@href}%
\providecommand \@href[1]{\@@startlink{#1}\@@href}%
\providecommand \@@href[1]{\endgroup#1\@@endlink}%
\providecommand \@sanitize@url [0]{\catcode `\\12\catcode `\$12\catcode
  `\&12\catcode `\#12\catcode `\^12\catcode `\_12\catcode `\%12\relax}%
\providecommand \@@startlink[1]{}%
\providecommand \@@endlink[0]{}%
\providecommand \url  [0]{\begingroup\@sanitize@url \@url }%
\providecommand \@url [1]{\endgroup\@href {#1}{\urlprefix }}%
\providecommand \urlprefix  [0]{URL }%
\providecommand \Eprint [0]{\href }%
\providecommand \doibase [0]{http://dx.doi.org/}%
\providecommand \selectlanguage [0]{\@gobble}%
\providecommand \bibinfo  [0]{\@secondoftwo}%
\providecommand \bibfield  [0]{\@secondoftwo}%
\providecommand \translation [1]{[#1]}%
\providecommand \BibitemOpen [0]{}%
\providecommand \bibitemStop [0]{}%
\providecommand \bibitemNoStop [0]{.\EOS\space}%
\providecommand \EOS [0]{\spacefactor3000\relax}%
\providecommand \BibitemShut  [1]{\csname bibitem#1\endcsname}%
\let\auto@bib@innerbib\@empty
\bibitem [{\citenamefont {Kibble}(1976)}]{Kibble}%
  \BibitemOpen
  \bibfield  {author} {\bibinfo {author} {\bibfnamefont {T.~W.~B.}\
  \bibnamefont {Kibble}},\ }\href {http://stacks.iop.org/0305-4470/9/i=8/a=029}
  {\bibfield  {journal} {\bibinfo  {journal} {Journal of Physics A:
  Mathematical and General}\ }\textbf {\bibinfo {volume} {9}},\ \bibinfo
  {pages} {1387} (\bibinfo {year} {1976})}\BibitemShut {NoStop}%
\bibitem [{\citenamefont {Kibble}(2004)}]{KibbleCite}%
  \BibitemOpen
  \bibfield  {author} {\bibinfo {author} {\bibfnamefont {T.}~\bibnamefont
  {Kibble}},\ }\href@noop {} {\bibfield  {journal} {\bibinfo  {journal}
  {COSLAB}\ } (\bibinfo {year} {2004})},\ \Eprint
  {http://arxiv.org/abs/astro-ph/0410073v2} {arXiv:astro-ph/0410073v2
  [astro-ph]} \BibitemShut {NoStop}%
\bibitem [{\citenamefont {Burgess}\ \emph {et~al.}(2001)\citenamefont
  {Burgess}, \citenamefont {Majumdar}, \citenamefont {Nolte}, \citenamefont
  {Quevedo}, \citenamefont {Rajesh},\ and\ \citenamefont
  {Zhang}}]{BurgessMajumdarNolteQuevedoRajeshZhang}%
  \BibitemOpen
  \bibfield  {author} {\bibinfo {author} {\bibfnamefont {C.~P.}\ \bibnamefont
  {Burgess}}, \bibinfo {author} {\bibfnamefont {M.}~\bibnamefont {Majumdar}},
  \bibinfo {author} {\bibfnamefont {D.}~\bibnamefont {Nolte}}, \bibinfo
  {author} {\bibfnamefont {F.}~\bibnamefont {Quevedo}}, \bibinfo {author}
  {\bibfnamefont {G.}~\bibnamefont {Rajesh}}, \ and\ \bibinfo {author}
  {\bibfnamefont {R.-J.}\ \bibnamefont {Zhang}},\ }\href {\doibase
  10.1088/1126-6708/2001/07/047} {\bibfield  {journal} {\bibinfo  {journal}
  {Journal of High Energy Physics}\ }\textbf {\bibinfo {volume} {2001}},\
  \bibinfo {pages} {047} (\bibinfo {year} {2001})},\ \Eprint
  {http://arxiv.org/abs/hep-th/0105204v3} {arXiv:hep-th/0105204v3 [hep-th]}
  \BibitemShut {NoStop}%
\bibitem [{\citenamefont {Dvali}\ \emph {et~al.}(2004)\citenamefont {Dvali},
  \citenamefont {Kallosh},\ and\ \citenamefont
  {Van~Proeyen}}]{DvaliKalloshProeyen}%
  \BibitemOpen
  \bibfield  {author} {\bibinfo {author} {\bibfnamefont {G.}~\bibnamefont
  {Dvali}}, \bibinfo {author} {\bibfnamefont {R.}~\bibnamefont {Kallosh}}, \
  and\ \bibinfo {author} {\bibfnamefont {A.}~\bibnamefont {Van~Proeyen}},\
  }\href {http://stacks.iop.org/1126-6708/2004/i=01/a=035} {\bibfield
  {journal} {\bibinfo  {journal} {JHEP}\ }\textbf {\bibinfo {volume} {2004}},\
  \bibinfo {pages} {035} (\bibinfo {year} {2004})},\ \Eprint
  {http://arxiv.org/abs/hep-th/0312005v3} {arXiv:hep-th/0312005v3 [hep-th]}
  \BibitemShut {NoStop}%
\bibitem [{\citenamefont {Copeland}\ \emph {et~al.}(2004)\citenamefont
  {Copeland}, \citenamefont {Myers},\ and\ \citenamefont
  {Polchinski}}]{PolchinskiCopelandMyers}%
  \BibitemOpen
  \bibfield  {author} {\bibinfo {author} {\bibfnamefont {E.~J.}\ \bibnamefont
  {Copeland}}, \bibinfo {author} {\bibfnamefont {R.~C.}\ \bibnamefont {Myers}},
  \ and\ \bibinfo {author} {\bibfnamefont {J.}~\bibnamefont {Polchinski}},\
  }\href {\doibase 10.1088/1126-6708/2004/06/013} {\bibfield  {journal}
  {\bibinfo  {journal} {JHEP}\ }\textbf {\bibinfo {volume} {2004}},\ \bibinfo
  {pages} {013} (\bibinfo {year} {2004})},\ \Eprint
  {http://arxiv.org/abs/hep-th/0312067v5} {arXiv:hep-th/0312067v5 [hep-th]}
  \BibitemShut {NoStop}%
\bibitem [{\citenamefont {Sarangi}\ and\ \citenamefont
  {Tye}(2002)}]{SarangiTye}%
  \BibitemOpen
  \bibfield  {author} {\bibinfo {author} {\bibfnamefont {S.}~\bibnamefont
  {Sarangi}}\ and\ \bibinfo {author} {\bibfnamefont {S.-H.~H.}\ \bibnamefont
  {Tye}},\ }\href {\doibase 10.1016/S0370-2693(02)01824-5} {\bibfield
  {journal} {\bibinfo  {journal} {Phys.Lett.B}\ }\textbf {\bibinfo {volume}
  {536}},\ \bibinfo {pages} {185} (\bibinfo {year} {2002})},\ \Eprint
  {http://arxiv.org/abs/hep-th/0204074v1} {arXiv:hep-th/0204074v1 [hep-th]}
  \BibitemShut {NoStop}%
\bibitem [{\citenamefont {Firouzjahi}\ and\ \citenamefont
  {Tye}(2005)}]{FirouzjahiTye}%
  \BibitemOpen
  \bibfield  {author} {\bibinfo {author} {\bibfnamefont {H.}~\bibnamefont
  {Firouzjahi}}\ and\ \bibinfo {author} {\bibfnamefont {S.-H.~H.}\ \bibnamefont
  {Tye}},\ }\href {\doibase 10.1088/1475-7516/2005/03/009} {\bibfield
  {journal} {\bibinfo  {journal} {JCAP}\ }\textbf {\bibinfo {volume} {0503}},\
  \bibinfo {pages} {009} (\bibinfo {year} {2005})},\ \Eprint
  {http://arxiv.org/abs/hep-th/0501099v3} {arXiv:hep-th/0501099v3 [hep-th]}
  \BibitemShut {NoStop}%
\bibitem [{\citenamefont {Jones}\ \emph {et~al.}(2003)\citenamefont {Jones},
  \citenamefont {Stoica},\ and\ \citenamefont {Tye}}]{JonesStoicaTye}%
  \BibitemOpen
  \bibfield  {author} {\bibinfo {author} {\bibfnamefont {N.~T.}\ \bibnamefont
  {Jones}}, \bibinfo {author} {\bibfnamefont {H.}~\bibnamefont {Stoica}}, \
  and\ \bibinfo {author} {\bibfnamefont {S.-H.~H.}\ \bibnamefont {Tye}},\
  }\href {\doibase 10.1016/S0370-2693(03)00592-6} {\bibfield  {journal}
  {\bibinfo  {journal} {Phys.Lett.}\ }\textbf {\bibinfo {volume} {B}},\
  \bibinfo {pages} {6} (\bibinfo {year} {2003})},\ \Eprint
  {http://arxiv.org/abs/hep-th/0303269v1} {arXiv:hep-th/0303269v1 [hep-th]}
  \BibitemShut {NoStop}%
\bibitem [{\citenamefont {Jeannerot}\ \emph {et~al.}(2003)\citenamefont
  {Jeannerot}, \citenamefont {Rocher},\ and\ \citenamefont
  {Sakellariadou}}]{JeannerotRocherSakellariadou}%
  \BibitemOpen
  \bibfield  {author} {\bibinfo {author} {\bibfnamefont {R.}~\bibnamefont
  {Jeannerot}}, \bibinfo {author} {\bibfnamefont {J.}~\bibnamefont {Rocher}}, \
  and\ \bibinfo {author} {\bibfnamefont {M.}~\bibnamefont {Sakellariadou}},\
  }\href {\doibase 10.1103/PhysRevD.68.103514} {\bibfield  {journal} {\bibinfo
  {journal} {Phys.Rev.D}\ }\textbf {\bibinfo {volume} {68}},\ \bibinfo {pages}
  {103514} (\bibinfo {year} {2003})},\ \Eprint
  {http://arxiv.org/abs/hep-ph/0308134v1} {arXiv:hep-ph/0308134v1 [hep-ph]}
  \BibitemShut {NoStop}%
\bibitem [{\citenamefont {Cui}\ \emph {et~al.}(2008)\citenamefont {Cui},
  \citenamefont {Martin}, \citenamefont {Morrissey},\ and\ \citenamefont
  {Wells}}]{CuiMartinMorrisseyWells}%
  \BibitemOpen
  \bibfield  {author} {\bibinfo {author} {\bibfnamefont {Y.}~\bibnamefont
  {Cui}}, \bibinfo {author} {\bibfnamefont {S.~P.}\ \bibnamefont {Martin}},
  \bibinfo {author} {\bibfnamefont {D.~E.}\ \bibnamefont {Morrissey}}, \ and\
  \bibinfo {author} {\bibfnamefont {J.~D.}\ \bibnamefont {Wells}},\ }\href
  {\doibase 10.1103/PhysRevD.77.043528} {\bibfield  {journal} {\bibinfo
  {journal} {Phys.Rev.D}\ }\textbf {\bibinfo {volume} {77}},\ \bibinfo {pages}
  {043528} (\bibinfo {year} {2008})},\ \Eprint
  {http://arxiv.org/abs/0709.0950v2} {arXiv:0709.0950v2 [hep-ph]} \BibitemShut
  {NoStop}%
\bibitem [{\citenamefont {Jeannerot}\ and\ \citenamefont
  {Postma}(2004)}]{JeannerotPostma}%
  \BibitemOpen
  \bibfield  {author} {\bibinfo {author} {\bibfnamefont {R.}~\bibnamefont
  {Jeannerot}}\ and\ \bibinfo {author} {\bibfnamefont {M.}~\bibnamefont
  {Postma}},\ }\href {\doibase 10.1088/1126-6708/2004/12/043} {\bibfield
  {journal} {\bibinfo  {journal} {JHEP}\ }\textbf {\bibinfo {volume} {0412}},\
  \bibinfo {pages} {043} (\bibinfo {year} {2004})},\ \Eprint
  {http://arxiv.org/abs/hep-ph/0411260} {arXiv:hep-ph/0411260 [astro-ph.CO]}
  \BibitemShut {NoStop}%
\bibitem [{\citenamefont {Ach\'ucarro}\ \emph {et~al.}(2006)\citenamefont
  {Ach\'ucarro}, \citenamefont {Celi}, \citenamefont {Esole}, \citenamefont
  {Van~den Bergh},\ and\ \citenamefont {A.}}]{AchucarroCeli}%
  \BibitemOpen
  \bibfield  {author} {\bibinfo {author} {\bibfnamefont {A.}~\bibnamefont
  {Ach\'ucarro}}, \bibinfo {author} {\bibfnamefont {A.}~\bibnamefont {Celi}},
  \bibinfo {author} {\bibfnamefont {M.}~\bibnamefont {Esole}}, \bibinfo
  {author} {\bibfnamefont {J.}~\bibnamefont {Van~den Bergh}}, \ and\ \bibinfo
  {author} {\bibfnamefont {V.~P.}\ \bibnamefont {A.}},\ }\href {\doibase
  10.1088/1126-6708/2006/01/102} {\bibfield  {journal} {\bibinfo  {journal}
  {JHEP}\ }\textbf {\bibinfo {volume} {0601}},\ \bibinfo {pages} {102}
  (\bibinfo {year} {2006})},\ \Eprint {http://arxiv.org/abs/hep-th/0511001v2}
  {arXiv:hep-th/0511001v2 [hep-th]} \BibitemShut {NoStop}%
\bibitem [{\citenamefont {Majumdar}\ and\ \citenamefont
  {Davis}(2002)}]{DavisMajumdar}%
  \BibitemOpen
  \bibfield  {author} {\bibinfo {author} {\bibfnamefont {M.}~\bibnamefont
  {Majumdar}}\ and\ \bibinfo {author} {\bibfnamefont {A.~C.}\ \bibnamefont
  {Davis}},\ }\href {http://stacks.iop.org/1126-6708/2002/i=03/a=056}
  {\bibfield  {journal} {\bibinfo  {journal} {JHEP}\ }\textbf {\bibinfo
  {volume} {2002}},\ \bibinfo {pages} {056} (\bibinfo {year} {2002})},\ \Eprint
  {http://arxiv.org/abs/hep-th/0202148v3} {arXiv:hep-th/0202148v3 [hep-th]}
  \BibitemShut {NoStop}%
\bibitem [{\citenamefont {Brandenberger}(2013)}]{BRANDENBERGER2}%
  \BibitemOpen
  \bibfield  {author} {\bibinfo {author} {\bibfnamefont {R.~H.}\ \bibnamefont
  {Brandenberger}},\ }\href@noop {} {\bibfield  {journal} {\bibinfo  {journal}
  {The Universe}\ }\textbf {\bibinfo {volume} {1}},\ \bibinfo {pages} {6 }
  (\bibinfo {year} {2013})},\ \Eprint {http://arxiv.org/abs/1401.4619v1}
  {arXiv:1401.4619v1 [astro-ph.CO]} \BibitemShut {NoStop}%
\bibitem [{\citenamefont {Kawasaki}\ \emph {et~al.}(2015)\citenamefont
  {Kawasaki}, \citenamefont {Saikawa},\ and\ \citenamefont
  {Sekiguchi}}]{KawasakiKen'ichiSekiguchi}%
  \BibitemOpen
  \bibfield  {author} {\bibinfo {author} {\bibfnamefont {M.}~\bibnamefont
  {Kawasaki}}, \bibinfo {author} {\bibfnamefont {K.}~\bibnamefont {Saikawa}}, \
  and\ \bibinfo {author} {\bibfnamefont {T.}~\bibnamefont {Sekiguchi}},\ }\href
  {\doibase 10.1103/PhysRevD.91.065014} {\bibfield  {journal} {\bibinfo
  {journal} {Phys. Rev. D}\ }\textbf {\bibinfo {volume} {91}},\ \bibinfo
  {pages} {065014} (\bibinfo {year} {2015})},\ \Eprint
  {http://arxiv.org/abs/1412.0789v3} {arXiv:1412.0789v3 [hep-ph]} \BibitemShut
  {NoStop}%
\bibitem [{\citenamefont {Gorghetto}\ \emph {et~al.}(2018)\citenamefont
  {Gorghetto}, \citenamefont {Hardy},\ and\ \citenamefont
  {Villadoro}}]{GorghettoHardyVilladoro}%
  \BibitemOpen
  \bibfield  {author} {\bibinfo {author} {\bibfnamefont {M.}~\bibnamefont
  {Gorghetto}}, \bibinfo {author} {\bibfnamefont {E.}~\bibnamefont {Hardy}}, \
  and\ \bibinfo {author} {\bibfnamefont {G.}~\bibnamefont {Villadoro}},\ }\href
  {\doibase 10.1007/JHEP07(2018)151} {\bibfield  {journal} {\bibinfo  {journal}
  {JHEP}\ }\textbf {\bibinfo {volume} {2018}} (\bibinfo {year} {2018}),\
  10.1007/JHEP07(2018)151},\ \Eprint {http://arxiv.org/abs/1806.04677}
  {arXiv:1806.04677 [hep-ph]} \BibitemShut {NoStop}%
\bibitem [{\citenamefont {Fleury}\ and\ \citenamefont
  {Moore}(2016)}]{FleuryMoore}%
  \BibitemOpen
  \bibfield  {author} {\bibinfo {author} {\bibfnamefont {L.}~\bibnamefont
  {Fleury}}\ and\ \bibinfo {author} {\bibfnamefont {G.~D.}\ \bibnamefont
  {Moore}},\ }\href {\doibase 10.1088/1475-7516/2016/01/004} {\bibfield
  {journal} {\bibinfo  {journal} {Journal of Cosmology and Astroparticle
  Physics}\ }\textbf {\bibinfo {volume} {2016}},\ \bibinfo {pages} {004}
  (\bibinfo {year} {2016})},\ \Eprint {http://arxiv.org/abs/1509.00026v1}
  {arXiv:1509.00026v1 [hep-ph]} \BibitemShut {NoStop}%
\bibitem [{\citenamefont {Gripaios}\ and\ \citenamefont
  {Randal-Williams}(2018)}]{GripaiosRandal-Williams}%
  \BibitemOpen
  \bibfield  {author} {\bibinfo {author} {\bibfnamefont {B.}~\bibnamefont
  {Gripaios}}\ and\ \bibinfo {author} {\bibfnamefont {O.}~\bibnamefont
  {Randal-Williams}},\ }\href {\doibase
  https://doi.org/10.1016/j.physletb.2018.05.013} {\bibfield  {journal}
  {\bibinfo  {journal} {Physics Letters B}\ }\textbf {\bibinfo {volume}
  {782}},\ \bibinfo {pages} {94 } (\bibinfo {year} {2018})},\ \Eprint
  {http://arxiv.org/abs/1610.05623v2} {arXiv:1610.05623v2 [hep-th]}
  \BibitemShut {NoStop}%
\bibitem [{\citenamefont {Spergel}\ and\ \citenamefont {Pen}(1997)}]{Spergel}%
  \BibitemOpen
  \bibfield  {author} {\bibinfo {author} {\bibfnamefont {D.}~\bibnamefont
  {Spergel}}\ and\ \bibinfo {author} {\bibfnamefont {U.-L.}\ \bibnamefont
  {Pen}},\ }\href {\doibase 10.1086/311074} {\bibfield  {journal} {\bibinfo
  {journal} {The Astrophysical Journal}\ }\textbf {\bibinfo {volume} {491}},\
  \bibinfo {pages} {L67} (\bibinfo {year} {1997})},\ \Eprint
  {http://arxiv.org/abs/astro-ph/9611198} {arXiv:astro-ph/9611198 [astro-ph]}
  \BibitemShut {NoStop}%
\bibitem [{\citenamefont {Jacobs}\ and\ \citenamefont
  {Rebbi}(1979)}]{JacobsRebbi}%
  \BibitemOpen
  \bibfield  {author} {\bibinfo {author} {\bibfnamefont {L.}~\bibnamefont
  {Jacobs}}\ and\ \bibinfo {author} {\bibfnamefont {C.}~\bibnamefont {Rebbi}},\
  }\href {\doibase 10.1103/PhysRevB.19.4486} {\bibfield  {journal} {\bibinfo
  {journal} {Phys. Rev. B}\ }\textbf {\bibinfo {volume} {19}},\ \bibinfo
  {pages} {4486} (\bibinfo {year} {1979})}\BibitemShut {NoStop}%
\bibitem [{\citenamefont {Saffin}(2005)}]{Saffin}%
  \BibitemOpen
  \bibfield  {author} {\bibinfo {author} {\bibfnamefont {P.~M.}\ \bibnamefont
  {Saffin}},\ }\href {\doibase 10.1088/1126-6708/2005/09/011} {\bibfield
  {journal} {\bibinfo  {journal} {Journal of High Energy Physics}\ }\textbf
  {\bibinfo {volume} {2005}},\ \bibinfo {pages} {011} (\bibinfo {year}
  {2005})},\ \Eprint {http://arxiv.org/abs/hep-th/0506138}
  {arXiv:hep-th/0506138 [hep-th]} \BibitemShut {NoStop}%
\bibitem [{\citenamefont {Copeland}\ \emph {et~al.}(2006)\citenamefont
  {Copeland}, \citenamefont {Kibble},\ and\ \citenamefont
  {Steer}}]{CopelandKibbleSteer}%
  \BibitemOpen
  \bibfield  {author} {\bibinfo {author} {\bibfnamefont {E.~J.}\ \bibnamefont
  {Copeland}}, \bibinfo {author} {\bibfnamefont {T.~W.~B.}\ \bibnamefont
  {Kibble}}, \ and\ \bibinfo {author} {\bibfnamefont {D.~A.}\ \bibnamefont
  {Steer}},\ }\href {\doibase 10.1103/PhysRevLett.97.021602} {\bibfield
  {journal} {\bibinfo  {journal} {Phys. Rev. Lett.}\ }\textbf {\bibinfo
  {volume} {97}},\ \bibinfo {pages} {021602} (\bibinfo {year} {2006})},\
  \Eprint {http://arxiv.org/abs/hep-th/0601153v3} {arXiv:hep-th/0601153v3
  [hep-th]} \BibitemShut {NoStop}%
\bibitem [{\citenamefont {Copeland}\ \emph {et~al.}(2007)\citenamefont
  {Copeland}, \citenamefont {Kibble},\ and\ \citenamefont
  {Steer}}]{CopelandKibbleSteer2}%
  \BibitemOpen
  \bibfield  {author} {\bibinfo {author} {\bibfnamefont {E.~J.}\ \bibnamefont
  {Copeland}}, \bibinfo {author} {\bibfnamefont {T.~W.~B.}\ \bibnamefont
  {Kibble}}, \ and\ \bibinfo {author} {\bibfnamefont {D.~A.}\ \bibnamefont
  {Steer}},\ }\href {\doibase 10.1103/PhysRevD.75.065024} {\bibfield  {journal}
  {\bibinfo  {journal} {Phys. Rev. D}\ }\textbf {\bibinfo {volume} {75}},\
  \bibinfo {pages} {065024} (\bibinfo {year} {2007})},\ \Eprint
  {http://arxiv.org/abs/hep-th/0611243v2} {arXiv:hep-th/0611243v2 [hep-th]}
  \BibitemShut {NoStop}%
\bibitem [{\citenamefont {Copeland}\ \emph {et~al.}(2008)\citenamefont
  {Copeland}, \citenamefont {Firouzjahi}, \citenamefont {Kibble},\ and\
  \citenamefont {Steer}}]{CopelandFirouzjahiKibbleSteer}%
  \BibitemOpen
  \bibfield  {author} {\bibinfo {author} {\bibfnamefont {E.~J.}\ \bibnamefont
  {Copeland}}, \bibinfo {author} {\bibfnamefont {H.}~\bibnamefont
  {Firouzjahi}}, \bibinfo {author} {\bibfnamefont {T.~W.~B.}\ \bibnamefont
  {Kibble}}, \ and\ \bibinfo {author} {\bibfnamefont {D.~A.}\ \bibnamefont
  {Steer}},\ }\href {\doibase 10.1103/PhysRevD.77.063521} {\bibfield  {journal}
  {\bibinfo  {journal} {Phys. Rev. D}\ }\textbf {\bibinfo {volume} {77}},\
  \bibinfo {pages} {063521} (\bibinfo {year} {2008})},\ \Eprint
  {http://arxiv.org/abs/0712.0808v1} {arXiv:0712.0808v1 [hep-th]} \BibitemShut
  {NoStop}%
\bibitem [{\citenamefont {Salmi}\ \emph {et~al.}(2008)\citenamefont {Salmi},
  \citenamefont {Ach\'ucarro}, \citenamefont {Copeland}, \citenamefont
  {Kibble}, \citenamefont {de~Putter},\ and\ \citenamefont
  {Steer}}]{SalmiAchucarroCopelandKibblePutterSteer}%
  \BibitemOpen
  \bibfield  {author} {\bibinfo {author} {\bibfnamefont {P.}~\bibnamefont
  {Salmi}}, \bibinfo {author} {\bibfnamefont {A.}~\bibnamefont {Ach\'ucarro}},
  \bibinfo {author} {\bibfnamefont {E.~J.}\ \bibnamefont {Copeland}}, \bibinfo
  {author} {\bibfnamefont {T.~W.~B.}\ \bibnamefont {Kibble}}, \bibinfo {author}
  {\bibfnamefont {R.}~\bibnamefont {de~Putter}}, \ and\ \bibinfo {author}
  {\bibfnamefont {D.~A.}\ \bibnamefont {Steer}},\ }\href {\doibase
  10.1103/PhysRevD.77.041701} {\bibfield  {journal} {\bibinfo  {journal} {Phys.
  Rev. D}\ }\textbf {\bibinfo {volume} {77}},\ \bibinfo {pages} {041701(R)}
  (\bibinfo {year} {2008})},\ \Eprint {http://arxiv.org/abs/0712.1204v2}
  {arXiv:0712.1204v2 [hep-th]} \BibitemShut {NoStop}%
\bibitem [{\citenamefont {Bevis}\ and\ \citenamefont
  {Saffin}(2008)}]{BevisSaffin}%
  \BibitemOpen
  \bibfield  {author} {\bibinfo {author} {\bibfnamefont {N.}~\bibnamefont
  {Bevis}}\ and\ \bibinfo {author} {\bibfnamefont {P.~M.}\ \bibnamefont
  {Saffin}},\ }\href {\doibase 10.1103/PhysRevD.78.023503} {\bibfield
  {journal} {\bibinfo  {journal} {Phys. Rev. D}\ }\textbf {\bibinfo {volume}
  {78}},\ \bibinfo {pages} {023503} (\bibinfo {year} {2008})},\ \Eprint
  {http://arxiv.org/abs/0804.0200v2} {arXiv:0804.0200v2 [hep-th]} \BibitemShut
  {NoStop}%
\bibitem [{\citenamefont {Avgoustidis}\ and\ \citenamefont
  {Shellard}(2008)}]{AvgoustidisShellard}%
  \BibitemOpen
  \bibfield  {author} {\bibinfo {author} {\bibfnamefont {A.}~\bibnamefont
  {Avgoustidis}}\ and\ \bibinfo {author} {\bibfnamefont {E.~P.~S.}\
  \bibnamefont {Shellard}},\ }\href {\doibase 10.1103/PhysRevD.78.103510}
  {\bibfield  {journal} {\bibinfo  {journal} {Phys. Rev. D}\ }\textbf {\bibinfo
  {volume} {78}},\ \bibinfo {pages} {103510} (\bibinfo {year} {2008})},\
  \Eprint {http://arxiv.org/abs/0705.3395v3} {arXiv:0705.3395v3 [astro-ph.CO]}
  \BibitemShut {NoStop}%
\bibitem [{\citenamefont {Rybak}\ \emph {et~al.}(2019)\citenamefont {Rybak},
  \citenamefont {Avgoustidis},\ and\ \citenamefont
  {Martins}}]{RybakAvgoustidisMartins3}%
  \BibitemOpen
  \bibfield  {author} {\bibinfo {author} {\bibfnamefont {I.~Y.}\ \bibnamefont
  {Rybak}}, \bibinfo {author} {\bibfnamefont {A.}~\bibnamefont {Avgoustidis}},
  \ and\ \bibinfo {author} {\bibfnamefont {C.~J. A.~P.}\ \bibnamefont
  {Martins}},\ }\href {\doibase 10.1103/PhysRevD.99.063516} {\bibfield
  {journal} {\bibinfo  {journal} {Phys. Rev. D}\ }\textbf {\bibinfo {volume}
  {99}},\ \bibinfo {pages} {063516} (\bibinfo {year} {2019})},\ \Eprint
  {http://arxiv.org/abs/1812.04584v2} {arXiv:1812.04584v2 [astro-ph.CO]}
  \BibitemShut {NoStop}%
\bibitem [{\citenamefont {Witten}(1985)}]{Witten}%
  \BibitemOpen
  \bibfield  {author} {\bibinfo {author} {\bibfnamefont {E.}~\bibnamefont
  {Witten}},\ }\href {\doibase 10.1016/0550-3213(85)90022-7} {\bibfield
  {journal} {\bibinfo  {journal} {Nuclear Physics B}\ }\textbf {\bibinfo
  {volume} {249}},\ \bibinfo {pages} {557 } (\bibinfo {year}
  {1985})}\BibitemShut {NoStop}%
\bibitem [{\citenamefont {Davis}\ \emph {et~al.}(2005)\citenamefont {Davis},
  \citenamefont {Binétruy},\ and\ \citenamefont
  {Davis}}]{DavisBinétruyDavis}%
  \BibitemOpen
  \bibfield  {author} {\bibinfo {author} {\bibfnamefont {S.~C.}\ \bibnamefont
  {Davis}}, \bibinfo {author} {\bibfnamefont {P.}~\bibnamefont {Binétruy}}, \
  and\ \bibinfo {author} {\bibfnamefont {A.-C.}\ \bibnamefont {Davis}},\ }\href
  {\doibase https://doi.org/10.1016/j.physletb.2005.02.038} {\bibfield
  {journal} {\bibinfo  {journal} {Physics Letters B}\ }\textbf {\bibinfo
  {volume} {611}},\ \bibinfo {pages} {39 } (\bibinfo {year} {2005})},\ \Eprint
  {http://arxiv.org/abs/hep-th/0501200v1} {arXiv:hep-th/0501200v1 [hep-th]}
  \BibitemShut {NoStop}%
\bibitem [{\citenamefont {Davis}\ \emph {et~al.}(1997)\citenamefont {Davis},
  \citenamefont {Davis},\ and\ \citenamefont {Trodden}}]{DavisDavisTrodden}%
  \BibitemOpen
  \bibfield  {author} {\bibinfo {author} {\bibfnamefont {S.~C.}\ \bibnamefont
  {Davis}}, \bibinfo {author} {\bibfnamefont {A.~C.}\ \bibnamefont {Davis}}, \
  and\ \bibinfo {author} {\bibfnamefont {M.}~\bibnamefont {Trodden}},\ }\href
  {\doibase 10.1016/S0370-2693(97)00642-4} {\bibfield  {journal} {\bibinfo
  {journal} {Phys.Lett.}\ }\textbf {\bibinfo {volume} {B}},\ \bibinfo {pages}
  {257} (\bibinfo {year} {1997})},\ \Eprint
  {http://arxiv.org/abs/hep-ph/9702360v1} {arXiv:hep-ph/9702360v1 [hep-ph]}
  \BibitemShut {NoStop}%
\bibitem [{\citenamefont {Davis}\ \emph {et~al.}(1998)\citenamefont {Davis},
  \citenamefont {Davis},\ and\ \citenamefont {Trodden}}]{DavisDavisTrodden2}%
  \BibitemOpen
  \bibfield  {author} {\bibinfo {author} {\bibfnamefont {S.~C.}\ \bibnamefont
  {Davis}}, \bibinfo {author} {\bibfnamefont {A.-C.}\ \bibnamefont {Davis}}, \
  and\ \bibinfo {author} {\bibfnamefont {M.}~\bibnamefont {Trodden}},\ }\href
  {\doibase 10.1103/PhysRevD.57.5184} {\bibfield  {journal} {\bibinfo
  {journal} {Phys. Rev. D}\ }\textbf {\bibinfo {volume} {57}},\ \bibinfo
  {pages} {5184} (\bibinfo {year} {1998})},\ \Eprint
  {http://arxiv.org/abs/hep-ph/9711313v1} {arXiv:hep-ph/9711313v1 [hep-ph]}
  \BibitemShut {NoStop}%
\bibitem [{\citenamefont {Allys}(2016)}]{Allys}%
  \BibitemOpen
  \bibfield  {author} {\bibinfo {author} {\bibfnamefont {E.}~\bibnamefont
  {Allys}},\ }\href {\doibase 10.1088/1475-7516/2016/04/009} {\bibfield
  {journal} {\bibinfo  {journal} {JCAP}\ }\textbf {\bibinfo {volume} {1604}},\
  \bibinfo {pages} {009} (\bibinfo {year} {2016})},\ \Eprint
  {http://arxiv.org/abs/1505.07888v3} {arXiv:1505.07888v3 [gr-qc]} \BibitemShut
  {NoStop}%
\bibitem [{\citenamefont {Sakellariadou}(2009)}]{Sakellariadou}%
  \BibitemOpen
  \bibfield  {author} {\bibinfo {author} {\bibfnamefont {M.}~\bibnamefont
  {Sakellariadou}},\ }\href {\doibase
  https://doi.org/10.1016/j.nuclphysbps.2009.07.046} {\bibfield  {journal}
  {\bibinfo  {journal} {Nuclear Physics B - Proceedings Supplements}\ }\textbf
  {\bibinfo {volume} {192-193}},\ \bibinfo {pages} {68 } (\bibinfo {year}
  {2009})},\ \bibinfo {note} {theory and Particle Physics: The LHC Perspective
  and Beyond},\ \Eprint {http://arxiv.org/abs/0902.0569v2} {arXiv:0902.0569v2
  [hep-th]} \BibitemShut {NoStop}%
\bibitem [{\citenamefont {Everett}(1988)}]{Everett}%
  \BibitemOpen
  \bibfield  {author} {\bibinfo {author} {\bibfnamefont {A.~E.}\ \bibnamefont
  {Everett}},\ }\href {\doibase 10.1103/PhysRevLett.61.1807} {\bibfield
  {journal} {\bibinfo  {journal} {Phys. Rev. Lett.}\ }\textbf {\bibinfo
  {volume} {61}},\ \bibinfo {pages} {1807} (\bibinfo {year}
  {1988})}\BibitemShut {NoStop}%
\bibitem [{\citenamefont {Hindmarsh}\ \emph {et~al.}(2016)\citenamefont
  {Hindmarsh}, \citenamefont {Rummukainen},\ and\ \citenamefont
  {Weir}}]{HindmarshRummukainenWeir}%
  \BibitemOpen
  \bibfield  {author} {\bibinfo {author} {\bibfnamefont {M.}~\bibnamefont
  {Hindmarsh}}, \bibinfo {author} {\bibfnamefont {K.}~\bibnamefont
  {Rummukainen}}, \ and\ \bibinfo {author} {\bibfnamefont {D.~J.}\ \bibnamefont
  {Weir}},\ }\href {\doibase 10.1103/PhysRevLett.117.251601} {\bibfield
  {journal} {\bibinfo  {journal} {Phys.Rev.Lett.}\ }\textbf {\bibinfo {volume}
  {117}},\ \bibinfo {pages} {251601} (\bibinfo {year} {2016})},\ \Eprint
  {http://arxiv.org/abs/arXiv:1607.00764v2} {arXiv:arXiv:1607.00764v2 [hep-th]}
  \BibitemShut {NoStop}%
\bibitem [{\citenamefont {Carter}(1989)}]{CARTER1989}%
  \BibitemOpen
  \bibfield  {author} {\bibinfo {author} {\bibfnamefont {B.}~\bibnamefont
  {Carter}},\ }\href {\doibase https://doi.org/10.1016/0370-2693(89)90976-3}
  {\bibfield  {journal} {\bibinfo  {journal} {Physics Letters B}\ }\textbf
  {\bibinfo {volume} {228}},\ \bibinfo {pages} {466 } (\bibinfo {year}
  {1989})}\BibitemShut {NoStop}%
\bibitem [{\citenamefont {Peter}(1992)}]{Peter1992}%
  \BibitemOpen
  \bibfield  {author} {\bibinfo {author} {\bibfnamefont {P.}~\bibnamefont
  {Peter}},\ }\href {\doibase 10.1103/PhysRevD.45.1091} {\bibfield  {journal}
  {\bibinfo  {journal} {Phys. Rev. D}\ }\textbf {\bibinfo {volume} {45}},\
  \bibinfo {pages} {1091} (\bibinfo {year} {1992})}\BibitemShut {NoStop}%
\bibitem [{\citenamefont {Carter}\ and\ \citenamefont
  {Peter}(1995)}]{CarterPeter}%
  \BibitemOpen
  \bibfield  {author} {\bibinfo {author} {\bibfnamefont {B.}~\bibnamefont
  {Carter}}\ and\ \bibinfo {author} {\bibfnamefont {P.}~\bibnamefont {Peter}},\
  }\href {\doibase 10.1103/PhysRevD.52.R1744} {\bibfield  {journal} {\bibinfo
  {journal} {Phys. Rev. D}\ }\textbf {\bibinfo {volume} {52}},\ \bibinfo
  {pages} {R1744} (\bibinfo {year} {1995})},\ \Eprint
  {http://arxiv.org/abs/hep-ph/9411425v1} {arXiv:hep-ph/9411425v1 [hep-ph]}
  \BibitemShut {NoStop}%
\bibitem [{\citenamefont {Carter}\ and\ \citenamefont
  {Peter}(1999)}]{CarterPeter2}%
  \BibitemOpen
  \bibfield  {author} {\bibinfo {author} {\bibfnamefont {B.}~\bibnamefont
  {Carter}}\ and\ \bibinfo {author} {\bibfnamefont {P.}~\bibnamefont {Peter}},\
  }\href {\doibase 10.1016/S0370-2693(99)01070-9} {\bibfield  {journal}
  {\bibinfo  {journal} {Phys.Lett.}\ }\textbf {\bibinfo {volume} {B}},\
  \bibinfo {pages} {41} (\bibinfo {year} {1999})},\ \Eprint
  {http://arxiv.org/abs/hep-th/9905025v1} {arXiv:hep-th/9905025v1 [hep-th]}
  \BibitemShut {NoStop}%
\bibitem [{\citenamefont {Carter}(2000)}]{Carter2000}%
  \BibitemOpen
  \bibfield  {author} {\bibinfo {author} {\bibfnamefont {B.}~\bibnamefont
  {Carter}},\ }\href {\doibase
  10.1002/(SICI)1521-3889(200005)9:3/5<247::AID-ANDP247>3.0.CO;2-5} {\bibfield
  {journal} {\bibinfo  {journal} {Ann. Phys.}\ }\textbf {\bibinfo {volume}
  {9}},\ \bibinfo {pages} {247} (\bibinfo {year} {2000})},\ \Eprint
  {http://arxiv.org/abs/hep-th/0002162v1} {arXiv:hep-th/0002162v1 [hep-th]}
  \BibitemShut {NoStop}%
\bibitem [{\citenamefont {Nielsen}(1980)}]{Nielsen}%
  \BibitemOpen
  \bibfield  {author} {\bibinfo {author} {\bibfnamefont {N.}~\bibnamefont
  {Nielsen}},\ }\href {\doibase https://doi.org/10.1016/0550-3213(80)90130-3}
  {\bibfield  {journal} {\bibinfo  {journal} {Nuclear Physics B}\ }\textbf
  {\bibinfo {volume} {167}},\ \bibinfo {pages} {249 } (\bibinfo {year}
  {1980})}\BibitemShut {NoStop}%
\bibitem [{\citenamefont {Nielsen}\ and\ \citenamefont
  {Olesen}(1987)}]{NielsenOlesen}%
  \BibitemOpen
  \bibfield  {author} {\bibinfo {author} {\bibfnamefont {N.}~\bibnamefont
  {Nielsen}}\ and\ \bibinfo {author} {\bibfnamefont {P.}~\bibnamefont
  {Olesen}},\ }\href {\doibase https://doi.org/10.1016/0550-3213(87)90498-6}
  {\bibfield  {journal} {\bibinfo  {journal} {Nuclear Physics B}\ }\textbf
  {\bibinfo {volume} {291}},\ \bibinfo {pages} {829 } (\bibinfo {year}
  {1987})}\BibitemShut {NoStop}%
\bibitem [{\citenamefont {Carter}(1990)}]{Carter90}%
  \BibitemOpen
  \bibfield  {author} {\bibinfo {author} {\bibfnamefont {B.}~\bibnamefont
  {Carter}},\ }\href {\doibase 10.1103/PhysRevD.41.3869} {\bibfield  {journal}
  {\bibinfo  {journal} {Phys.Rev.}\ }\textbf {\bibinfo {volume} {D41}},\
  \bibinfo {pages} {3869} (\bibinfo {year} {1990})}\BibitemShut {NoStop}%
\bibitem [{\citenamefont {Vilenkin}(1990)}]{Vilenkin}%
  \BibitemOpen
  \bibfield  {author} {\bibinfo {author} {\bibfnamefont {A.}~\bibnamefont
  {Vilenkin}},\ }\href {\doibase 10.1103/PhysRevD.41.3038} {\bibfield
  {journal} {\bibinfo  {journal} {Phys. Rev.}\ }\textbf {\bibinfo {volume}
  {D41}},\ \bibinfo {pages} {3038} (\bibinfo {year} {1990})}\BibitemShut
  {NoStop}%
\bibitem [{\citenamefont {Martin}(1995)}]{Martin}%
  \BibitemOpen
  \bibfield  {author} {\bibinfo {author} {\bibfnamefont {X.}~\bibnamefont
  {Martin}},\ }\href {\doibase 10.1103/PhysRevLett.74.3102} {\bibfield
  {journal} {\bibinfo  {journal} {Phys. Rev. Lett.}\ }\textbf {\bibinfo
  {volume} {74}},\ \bibinfo {pages} {3102} (\bibinfo {year}
  {1995})}\BibitemShut {NoStop}%
\bibitem [{\citenamefont {Carter}(1995)}]{Carter95}%
  \BibitemOpen
  \bibfield  {author} {\bibinfo {author} {\bibfnamefont {B.}~\bibnamefont
  {Carter}},\ }\href {\doibase 10.1103/PhysRevLett.74.3098} {\bibfield
  {journal} {\bibinfo  {journal} {Phys.Rev.Lett.}\ }\textbf {\bibinfo {volume}
  {74}},\ \bibinfo {pages} {3098} (\bibinfo {year} {1995})},\ \Eprint
  {http://arxiv.org/abs/hep-th/9411231v1} {arXiv:hep-th/9411231v1 [hep-th]}
  \BibitemShut {NoStop}%
\bibitem [{\citenamefont {Steer}\ \emph {et~al.}(2018)\citenamefont {Steer},
  \citenamefont {Lilley}, \citenamefont {Yamauchi},\ and\ \citenamefont
  {Hiramatsu}}]{SteerLilleyYamauchiHiramatsu}%
  \BibitemOpen
  \bibfield  {author} {\bibinfo {author} {\bibfnamefont {D.~A.}\ \bibnamefont
  {Steer}}, \bibinfo {author} {\bibfnamefont {M.}~\bibnamefont {Lilley}},
  \bibinfo {author} {\bibfnamefont {D.}~\bibnamefont {Yamauchi}}, \ and\
  \bibinfo {author} {\bibfnamefont {T.}~\bibnamefont {Hiramatsu}},\ }\href
  {\doibase 10.1103/PhysRevD.97.023507} {\bibfield  {journal} {\bibinfo
  {journal} {Phys. Rev. D}\ }\textbf {\bibinfo {volume} {97}},\ \bibinfo
  {pages} {023507} (\bibinfo {year} {2018})},\ \Eprint
  {http://arxiv.org/abs/1710.07475v1} {arXiv:1710.07475v1 [astro-ph.CO]}
  \BibitemShut {NoStop}%
\bibitem [{\citenamefont {Blanco-Pillado}\ \emph {et~al.}(2001)\citenamefont
  {Blanco-Pillado}, \citenamefont {Olum},\ and\ \citenamefont
  {Vilenkin}}]{Blanco-PilladoOlumVilenkin}%
  \BibitemOpen
  \bibfield  {author} {\bibinfo {author} {\bibfnamefont {J.~J.}\ \bibnamefont
  {Blanco-Pillado}}, \bibinfo {author} {\bibfnamefont {K.~D.}\ \bibnamefont
  {Olum}}, \ and\ \bibinfo {author} {\bibfnamefont {A.}~\bibnamefont
  {Vilenkin}},\ }\href {\doibase 10.1103/PhysRevD.63.103513} {\bibfield
  {journal} {\bibinfo  {journal} {Phys. Rev. D}\ }\textbf {\bibinfo {volume}
  {63}},\ \bibinfo {pages} {103513} (\bibinfo {year} {2001})},\ \Eprint
  {http://arxiv.org/abs/astro-ph/0004410v3} {arXiv:astro-ph/0004410v3
  [astro-ph]} \BibitemShut {NoStop}%
\bibitem [{\citenamefont {Carter}(1997)}]{Carter2}%
  \BibitemOpen
  \bibfield  {author} {\bibinfo {author} {\bibfnamefont {B.}~\bibnamefont
  {Carter}},\ }\href@noop {} {\bibfield  {journal} {\bibinfo  {journal} {2nd
  Mexican School on Gravitation and Mathematical Physics}\ } (\bibinfo {year}
  {1997})},\ \Eprint {http://arxiv.org/abs/hep-th/9705172}
  {arXiv:hep-th/9705172 [hep-th]} \BibitemShut {NoStop}%
\bibitem [{\citenamefont {Rybak}\ \emph {et~al.}(2017)\citenamefont {Rybak},
  \citenamefont {Avgoustidis},\ and\ \citenamefont
  {Martins}}]{RybakAvgoustidisMartins}%
  \BibitemOpen
  \bibfield  {author} {\bibinfo {author} {\bibfnamefont {I.~Y.}\ \bibnamefont
  {Rybak}}, \bibinfo {author} {\bibfnamefont {A.}~\bibnamefont {Avgoustidis}},
  \ and\ \bibinfo {author} {\bibfnamefont {C.~J. A.~P.}\ \bibnamefont
  {Martins}},\ }\href {\doibase 10.1103/PhysRevD.96.103535} {\bibfield
  {journal} {\bibinfo  {journal} {Phys. Rev. D}\ }\textbf {\bibinfo {volume}
  {96}},\ \bibinfo {pages} {103535} (\bibinfo {year} {2017})},\ \Eprint
  {http://arxiv.org/abs/1709.01839v2} {arXiv:1709.01839v2 [astro-ph.CO]}
  \BibitemShut {NoStop}%
\bibitem [{\citenamefont {Peter}(1993)}]{Peter1993}%
  \BibitemOpen
  \bibfield  {author} {\bibinfo {author} {\bibfnamefont {P.}~\bibnamefont
  {Peter}},\ }\href {\doibase 10.1103/PhysRevD.47.3169} {\bibfield  {journal}
  {\bibinfo  {journal} {Phys. Rev. D}\ }\textbf {\bibinfo {volume} {47}},\
  \bibinfo {pages} {3169} (\bibinfo {year} {1993})}\BibitemShut {NoStop}%
\bibitem [{\citenamefont {Davis}\ \emph {et~al.}(2000)\citenamefont {Davis},
  \citenamefont {Kibble}, \citenamefont {Pickles},\ and\ \citenamefont
  {Steer}}]{DavisKibblePicklesSteer}%
  \BibitemOpen
  \bibfield  {author} {\bibinfo {author} {\bibfnamefont {A.~C.}\ \bibnamefont
  {Davis}}, \bibinfo {author} {\bibfnamefont {T.~W.~B.}\ \bibnamefont
  {Kibble}}, \bibinfo {author} {\bibfnamefont {M.}~\bibnamefont {Pickles}}, \
  and\ \bibinfo {author} {\bibfnamefont {D.~A.}\ \bibnamefont {Steer}},\ }\href
  {\doibase 10.1103/PhysRevD.62.083516} {\bibfield  {journal} {\bibinfo
  {journal} {Phys. Rev. D}\ }\textbf {\bibinfo {volume} {62}},\ \bibinfo
  {pages} {083516} (\bibinfo {year} {2000})},\ \Eprint
  {http://arxiv.org/abs/astro-ph/0005514v1} {arXiv:astro-ph/0005514v1
  [astro-ph]} \BibitemShut {NoStop}%
\bibitem [{\citenamefont {Rybak}\ \emph {et~al.}(2018)\citenamefont {Rybak},
  \citenamefont {Avgoustidis},\ and\ \citenamefont
  {Martins}}]{RybakAvgoustidisMartins2}%
  \BibitemOpen
  \bibfield  {author} {\bibinfo {author} {\bibfnamefont {I.~Y.}\ \bibnamefont
  {Rybak}}, \bibinfo {author} {\bibfnamefont {A.}~\bibnamefont {Avgoustidis}},
  \ and\ \bibinfo {author} {\bibfnamefont {C.~J. A.~P.}\ \bibnamefont
  {Martins}},\ }\href {\doibase 10.1103/PhysRevD.98.063519} {\bibfield
  {journal} {\bibinfo  {journal} {Phys. Rev. D}\ }\textbf {\bibinfo {volume}
  {98}},\ \bibinfo {pages} {063519} (\bibinfo {year} {2018})},\ \Eprint
  {http://arxiv.org/abs/1809.04033v1} {arXiv:1809.04033v1 [astro-ph.CO]}
  \BibitemShut {NoStop}%
\bibitem [{\citenamefont {Carter}\ and\ \citenamefont
  {Steer}(2004)}]{CarterSteer}%
  \BibitemOpen
  \bibfield  {author} {\bibinfo {author} {\bibfnamefont {B.}~\bibnamefont
  {Carter}}\ and\ \bibinfo {author} {\bibfnamefont {D.~A.}\ \bibnamefont
  {Steer}},\ }\href {\doibase 10.1103/PhysRevD.69.125002} {\bibfield  {journal}
  {\bibinfo  {journal} {Phys. Rev. D}\ }\textbf {\bibinfo {volume} {69}},\
  \bibinfo {pages} {125002} (\bibinfo {year} {2004})},\ \Eprint
  {http://arxiv.org/abs/hep-th/0307161} {arXiv:hep-th/0307161 [hep-th]}
  \BibitemShut {NoStop}%
\bibitem [{\citenamefont {Dasgupta}\ and\ \citenamefont
  {Mukhi}(1998)}]{DasguptaMukhi}%
  \BibitemOpen
  \bibfield  {author} {\bibinfo {author} {\bibfnamefont {K.}~\bibnamefont
  {Dasgupta}}\ and\ \bibinfo {author} {\bibfnamefont {S.}~\bibnamefont
  {Mukhi}},\ }\href {\doibase https://doi.org/10.1016/S0370-2693(98)00140-3}
  {\bibfield  {journal} {\bibinfo  {journal} {Physics Letters B}\ }\textbf
  {\bibinfo {volume} {423}},\ \bibinfo {pages} {261 } (\bibinfo {year}
  {1998})},\ \Eprint {http://arxiv.org/abs/hep-th/9711094v1}
  {arXiv:hep-th/9711094v1 [hep-th]} \BibitemShut {NoStop}%
\end{thebibliography}%
\end{document}